\documentclass[lettersize,journal]{IEEEtran}
\usepackage{cite}
\usepackage{amsmath,amssymb,amsfonts}
\usepackage{graphicx}
\usepackage{textcomp}
\usepackage[nolist]{acronym}
\usepackage{longtable}
\usepackage{relsize}
\usepackage{threeparttable}
\usepackage[normalem]{ulem}
\usepackage{comment}
\usepackage{color,soul}
\usepackage{multicol,multirow,xcolor,colortbl}
\usepackage{makecell}
\usepackage{orcidlink}
\usepackage{xparse}
\usepackage{caption}
\usepackage{subcaption}
\usepackage{algorithm}
\usepackage[noend]{algpseudocode}
\usepackage{svg}
\usepackage{pgfplots}
\usepackage{pgfplotstable}

\begin{acronym} 
\acro{3GPP}{Third Generation Partnership Project}
\acro{5G}{fifth generation}
\acro{AGC}{automatic gain control}
\acro{AoA}{angle of arrival}
\acro{AoD}{angle of departure}
\acro{AWGN}{additive white Gaussian noise}
\acro{BKC}{backward cancellation}
\acro{BS}{base station}
\acro{C-V2X}{cellular-vehicle-to-everything}
\acro{CA}{carrier aggregation}
\acro{CAD}{cooperative automated driving}
\acro{CAM}{cooperative awareness message}
\acro{CAV}{connected and autonomous vehicle}
\acro{CBR}{channel busy ratio}
\acro{CCDF}{complementary cumulative distribution function}
\acro{CD}{collision detection}
\acro{CPM}{collective perception message}
\acro{CRLB}{Cramer-Rao Lower Bound }
\acro{CRC}{cyclic redundancy check}
\acro{CRDSA}{contention resolution diversity slotted ALOHA}
\acro{CSA}{coded-slotted ALOHA}
\acro{CSI}{channel state information}
\acro{D2D}{device-to-device}
\acro{DENM}{decentralized environmental notification message}
\acro{DL}{downlink}
\acro{DMRS}{demodulation reference signal}
\acro{EED}{end-to-end delay}
\acro{FD}{in-band full-duplex}
\acro{FIM}{Fisher information matrix}
\acro{FR1}{frequency range 1}
\acro{FR2}{frequency range 2}
\acro{FRC}{forward cancellation}
\acro{GNSS}{global navigation satellite system}
\acro{GPS}{global positioning system}
\acro{HD}{half-duplex}
\acro{IBE}{in-band emission}
\acro{IIoT}{industrial Internet of things}
\acro{ISAC}{integrated sensing and communication}
\acro{ITS}{intelligent transport system}
\acro{JCS}{joint communication and sensing}
\acro{LMS}{least mean square}
\acro{LOS}{line-of-sight}
\acro{LTE}{long term evolution}  
\acro{MAC}{medium access control}
\acro{MCS}{modulation and coding scheme}
\acro{MIMO}{multiple input multiple output}
\acro{MRC}{maximum ratio combining}
\acro{NLOS}{non-line-of-sight}
\acro{NOMA}{non-orthogonal multiple access}
\acro{NR}{new radio}
\acro{NR-V2X}{new radio-vehicle-to-everything}
\acro{OFDM}{orthogonal frequency-division multiplexing}
\acro{PCB}{probabilistic codebook}
\acro{PEB}{position error bound}
\acro{PHY}{physical}
\acro{PRB}{physical resource block}
\acro{PRR}{packet reception ratio}
\acro{QoS}{quality of service}
\acro{RB-NOMA}{repetition-based non-orthogonal multiple access}
\acro{RCS}{radar cross section}
\acro{RRI}{resource reservation interval}
\acro{RSRP}{reference signal received power}
\acro{RSRPP}{reference signal received path power}
\acro{RSRQ}{reference signal received quality}
\acro{RSTD}{reference signal time difference}
\acro{RSU}{road side unit}
\acro{RTOA}{relative time of arrival}
\acro{RTT}{round trip time}
\acro{RX}{receiver}
\acro{SB-DS}{sensing-based dynamic scheduling}
\acro{SB-SPS}{sensing-based semi-persistent scheduling}
\acro{SCI}{sidelink control information}
\acro{SCS}{subcarrier spacing}
\acro{SIC}{successive interference cancellation}
\acro{SINR}{signal-to-interference-plus-noise ratio}
\acro{SL}{sidelink}
\acro{SL PRS}{sidelink positioning reference signal}
\acro{SNR}{signal-to-noise ratio}
\acro{SS}{sychronization signal}
\acro{TB}{transport block}
\acro{TDD}{time-division duplex}
\acro{TDoA}{time difference of arrival}
\acro{TOA}{time of arrival}
\acro{TTI}{transmission time interval}
\acro{TX}{transmitter}
\acro{UE}{user equipment}
\acro{UCA}{uniform circular array}
\acro{ULA}{uniform linear array}
\acro{V2V}{vehicle-to-vehicle} 
\acro{V2X}{vehicle-to-everything} 
\acro{VUE}{vehicular user equipment}
\acro{WBS}{wireless blind spot}
\acro{WBSP}{wireless blind spot probability}
\end{acronym}

\newcommand{\Prcv}[1]{{{P_\text{r}}_{#1}}}

\newcommand{\Nbusy}{N_\text{busy}}
\newcommand{\Ncbr}{N_\text{CBR}}

\newcommand{\Pnoise}{{P_\text{n}}}

\newcommand{\Uset}{\mathcal{U}}

\newcommand{\kNOMA}{k_\text{N}}

\newcommand{\IBEcoeff}{{\eta}}

\newcommand{\SINRt}{\overline{\gamma}}

\newcommand{\pkeep}{{p_\text{k}}}

\begin{document}

\title{Exploiting Repetitions and Interference Cancellation for the 6G-V2X Sidelink Autonomous Mode}

\author{\IEEEauthorblockN{
Alessandro Bazzi, Vittorio Todisco, Antonella Molinaro,\\ 
Antoine O. Berthet, Richard A. Stirling-Gallacher, 
Claudia Campolo
}
\thanks{A. Bazzi and V. Todisco are with DEI, Universit\`a di Bologna, Italy. }\thanks{
A. Molinaro and C. Campolo are with Universit\`a Mediterranea di Reggio Calabria, Italy. }\thanks{
A. O. Berthet is with Universit\'e Paris-Saclay, CNRS, CentraleSup\'elec, Gif-sur-Yvette, France. }\thanks{
R. A. Stirling-Gallacher is with Munich Research Center, Huawei Technologies Duesseldorf GmbH, Germany. }\thanks{
A. Bazzi, V. Todisco, A. Molinaro, A. O. Berthet, and C. Campolo are also with CNIT/WiLab, Bologna, Italy.}
\thanks{Corresponding author: A. Bazzi (email: alessandro.bazzi@unibo.it).}
\thanks{Copyright (c) 2025 IEEE. Personal use of this material is permitted. However, permission to use this material for any other purposes must be obtained from the IEEE by sending a request to pubs-permissions@ieee.org.}
}

\markboth{IEEE Transactions on Vehicular Technology, accepted for publication}%
{Shell \MakeLowercase{\textit{et al.}}: A Sample Article Using IEEEtran.cls for IEEE Journals}

\maketitle

\begin{abstract}
In recent years, the Third Generation Partnership Project (3GPP) has developed the new radio-vehicle-to-everything (NR-V2X) sidelink standard, to enable direct communication between \acp{CAV}. Users can autonomously select radio resources for their transmissions with the Mode 2 channel access scheme, which can also operate under out-of-coverage conditions. However, Mode 2 performance is hindered by interference and packet collisions arising from dynamic mobile environments and limitations in assessing radio resource availability. The 3GPP specifications allow transmitting multiple copies of the same packet to improve reliability, though at the cost of increased channel congestion. This paper proposes to leverage receivers equipped with \ac{SIC} capabilities, to exploit packet repetitions. Specifically, once a packet is successfully decoded the interfering contribution carried by repetitions can be cancelled from future or past received signals, enabling the decoding of new packets. Extensive highway scenario simulations demonstrate that the proposed solution significantly outperforms the legacy Mode 2 scheme, especially under high interference conditions, achieving improvements exceeding 100\% in some cases.
\end{abstract}


\acresetall

\section{Introduction}

In recent years, the \ac{3GPP} has undertaken significant efforts to overhaul the architectural and protocol components of cellular networks to meet the stringent requirements of future \acp{CAV}. As part of the updated \ac{5G} specifications, referred to as \ac{NR-V2X} \cite{bazzi2021design,GarMolBobGozColSahKou:21}, considerable emphasis has been placed on the \textit{sidelink} design, which enables direct \ac{V2V} communication.

A distributed resource allocation scheme, known as Mode~2, has been introduced as an enhancement of the earlier Mode~4 from previous \ac{3GPP} releases. Mode~2 enables vehicles to autonomously select sidelink radio resources. Since this scheme does not require a connection to \acp{BS}, it can operate effectively in driving scenarios both within and beyond cellular network coverage.

In Mode~2, resources are selected to minimize concurrent transmissions with neighbouring vehicles. This process relies on channel sensing and control information embedded in each packet, which indicates future radio resource usage. However, inherent limitations of channel sensing and signalling procedures, such as their inability to fully capture the dynamics of the vehicular environment, combined with the presence of hidden terminals, prevent Mode~2 from completely mitigating interference and packet collisions.

Furthermore, a significant portion of failed transmissions cannot be detected by the sender because most messages rely on unacknowledged broadcast transmissions. This is particularly true for advanced CAV applications, such as cooperative awareness, collective perception, and coordinated maneuvering services \cite{bazzi2024mco,9762711,10195908}.
To address this issue, Mode~2 allows packets to be retransmitted multiple times on different sidelink radio resources through a mechanism known as \textit{blind} retransmissions. In this approach, when enabled, retransmissions are enforced regardless of whether the original packet has been successfully decoded. 
On the one hand, packet retransmissions increase the likelihood of successful decoding by leveraging both time and frequency diversity in the transmitted copies. On the other hand, they increase channel load and may cause further congestion, especially in high-density scenarios. 

In this work, we propose a shift in perspective by exploiting the interference generated by multiple transmissions, rather than viewing it as an obstacle. Specifically, we propose enabling \acp{VUE} to leverage advanced receivers equipped with \ac{SIC} capabilities in the presence of blind retransmissions.

The concept behind \ac{SIC} is that the receiver continuously stores incoming signals while attempting to detect and successfully decode a packet. Once a packet is correctly decoded, the originally transmitted signal is reconstructed and subtracted from the received signal. This process reveals residual signals, which may correspond to weaker transmissions of packets from other sources, allowing these packets to be detected and decoded as well. Furthermore, the decoded packets are used to cancel interference caused by other copies of the same packet transmitted in different time intervals, thereby enabling the successful decoding of new packets that may otherwise remain corrupted by interference. 

The principle of exploiting and managing interference through \ac{SIC}-enabled receivers, rather than avoiding it \textit{a priori}, is a defining characteristic of \ac{NOMA} approaches \cite{hirai2019,10078378,di2017,9287485,todisco2023sic}. Recently, this concept has been applied in NOMA-inspired random access protocols, such as \ac{CRDSA} \cite{casini2007} and \ac{CSA} \cite{liva2011,7302046}, by leveraging the presence of multiple copies of a packet. Building on these approaches, our proposal enhances Mode~2 performance by incorporating \ac{SIC} capabilities and exploiting packet repetitions. This approach is hereafter referred to as \textit{repetition-based NOMA (RB-NOMA)}\acused{RB-NOMA}. 

The proposed \ac{RB-NOMA} advances the state of the art by offering the following original contributions:

\begin{itemize} 
\item It overcomes the inherent trade-off in NR-V2X Mode~2 operations between reliability and channel load. By utilizing \ac{SIC}-enabled receivers, the scheme leverages retransmissions to enhance reliability while removing at the receiver the interference contribution of decoded signals;
\item It introduces two novel operations for \ac{SIC}-enabled receivers, which apply \ac{SIC} to both future (\textit{forward cancellation}) and past (\textit{backward cancellation}) packet repetitions by buffering, respectively, correctly received packets and signal IQ samples, and iteratively canceling each correctly decoded packet; 
\item The required modifications to the 3GPP specifications are minimal and result in negligible control overhead increase. 
\end{itemize}

Simulation results, conducted in a wide range of settings, including varying vehicle densities, number of retransmissions, and data generation patterns, demonstrate significant improvements over legacy, non-\ac{SIC}-based operations.

The remainder of the paper is organized as follows. Section~\ref{sec:background} reviews the fundamentals of NR-V2X Mode~2 and summarizes related work. Section~\ref{sec:receivers} discusses the proposed \ac{RB-NOMA} and the enhanced receiver operations, followed by their modeling in Section~\ref{sec:SICmodeling}. Simulation results are presented in Section~\ref{sec:results}, and conclusions are drawn in Section~\ref{sec:conclusions}.

\section{Background}\label{sec:background}

This section provides an overview of 3GPP specifications for Mode~2 in NR-V2X sidelink \cite{bazzi2021design,GarMolBobGozColSahKou:21} and reviews the key related work that inspired our proposal.

\subsection{NR-V2X Sidelink with Mode~2}

\acp{CAV} can communicate directly with each other via the sidelink. In Mode~2, specifically, \acp{VUE} autonomously select sidelink radio resources without requiring cellular coverage.

 \begin{figure}[t]
\centering
\includegraphics[width=1\columnwidth]{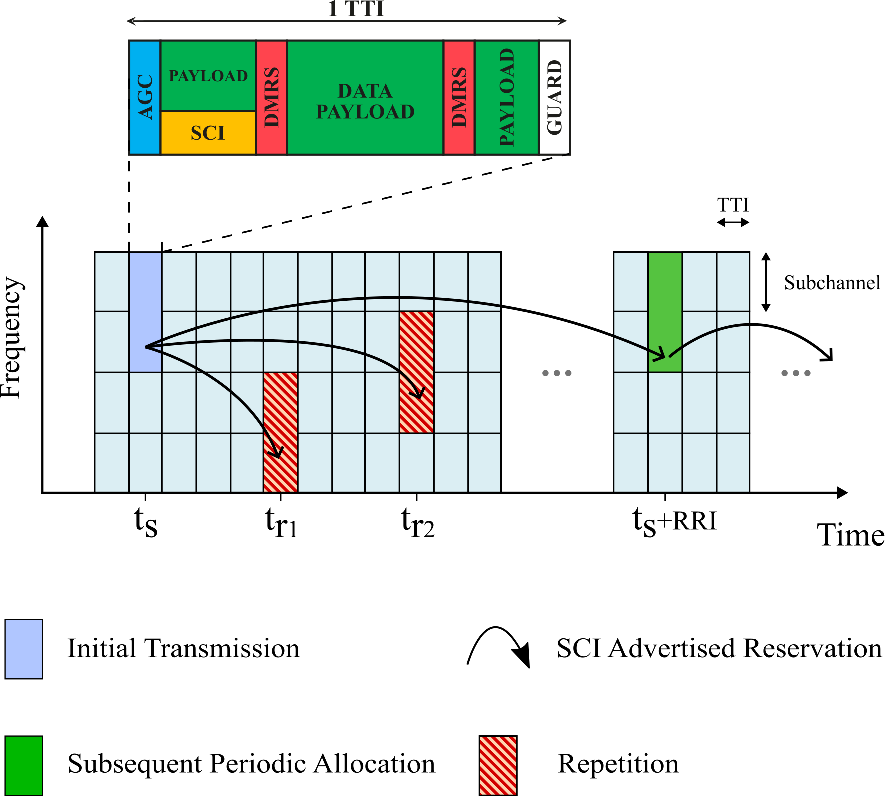}
\caption{Example of signal structure and representation of indications to following resource reservations.}
\label{fig:signal}
\end{figure}

Sidelink foresees \ac{OFDM} with a grid of time and frequency resources to which all \acp{VUE} are synchronized. In the time domain, a group of 14 \ac{OFDM} symbols constitutes a \textit{slot} or a \ac{TTI}, while in the frequency domain, a group of 12 subcarriers spaced by the \ac{SCS} forms a \ac{PRB}. \acp{PRB}, in turn, are grouped according to network configuration settings to form \textit{subchannels}. The duration of a \ac{TTI} and the bandwidth of \acp{PRB} depend on the so-called numerology, defined by a parameter $\mu$ that can take integer values between 0 and 3. Specifically, the \ac{TTI} lasts $2^{-\mu}$~ms, and the \ac{SCS} is $15\cdot2^\mu$~kHz.

Each packet is transmitted during a single TTI (slot), using one or multiple adjacent subchannels. The signal carries both control bits in the \ac{SCI}, \acp{DMRS} for channel estimation, and data bits in the \ac{TB}. The first OFDM symbol is a copy of the second symbol for detection and \ac{AGC}, while the final symbol of the \ac{TTI} is left empty to facilitate the switch from transmission to reception. 

The SCI carries essential control parameters, such as modulation, coding rate, and transmission format, that allow the receiving UE to correctly decode the associated data. If the SCI is not successfully decoded, the entire transmission cannot be recovered. Since sidelink communication may operate without centralized scheduling, the SCI allows devices to autonomously inform one another of their transmission configuration. Beyond basic decoding parameters, the SCI also includes two fields used to reserve resources for future transmissions \cite{38.212}: \textit{(i)} the \ac{RRI}, indicating a time offset at which the same subchannels will be reused, which is used to support periodic transmissions (with intervals that are usually in the range between 100~ms and 1~s); \textit{(ii)} the time resource assignment (TRA) and frequency resource assignment (FRA), which  allow the reservation of up to two transmission opportunities with full flexibility in both time and frequency, within a 32-TTI window from the current transmission; this mechanism is used to indicate the repetitions of the same packet. An example of signal structure is shown in Fig.~\ref{fig:signal}, together with the two types of resource reservations.

When users autonomously select their sidelink resources (i.e., subchannel and slot) according to Mode~2, each \ac{VUE} relies on the information decoded from received \acp{SCI} and on the measured signal power to identify and discard reserved resources. More specifically, the resource allocation follows either the \ac{SB-SPS} or the \ac{SB-DS}. The former is designed for periodic traffic, whereas the latter is better suited for aperiodic traffic. The main common aspects of these two schemes are as follows \cite{38.214}:

\begin{enumerate}
    \item A \ac{VUE} with a new packet to transmit considers two windows relative to the generation instant: a past \textit{sensing window}, used to retrieve information about allocated and reserved resources, and a future \textit{selection window}, where resources for the new transmission must be selected. The selection windows is set consistently with the maximum allowed delay, also called delay budget.

    \item The \ac{VUE} reads the \acp{SCI} of the packets received in the sensing window, which can carry information about future resource reservations. The \ac{VUE} only considers reservations associated with signals that have a received power level above a given threshold.

    \item The collected information is used to mark the announced resources as reserved (and thus unavailable) within the selection window.
    
    \item If the remaining available resources within the selection window are less than a given percentage, the \ac{VUE} returns to step 2 with an increased power threshold. This adjustment generally results in ignoring reservations made by farther vehicles. If necessary, the power threshold is increased iteratively until sufficient resources are identified.
    
    \item The \ac{VUE} randomly selects resources for its new transmission from the remaining available resources.
\end{enumerate}

In the case of \ac{SB-SPS}, once sidelink resources are selected for the current transmission, they can be reserved at regular intervals to support periodically generated data traffic. In this case, the \ac{SCI} reserves the same resources after a number of slots equal to the configured \ac{RRI}. Such reservations may last for a randomly selected number of successive transmissions, after which a new resource selection is performed following the same procedure. In contrast, with \ac{SB-DS}, a new resource selection is performed for each newly generated packet.

In both cases, a packet can be retransmitted up to 32 times to improve reliability, which is particularly useful for broadcast packets, as sidelink broadcast transmissions do not support acknowledgments. These retransmissions, also called replicas, are \textit{blind}, meaning they are delivered regardless of whether any of the replicas have been successfully received by neighboring vehicles.
When blind retransmissions are used, the \ac{VUE} randomly selects a distinct resource for the initial transmission and each replica. As discussed earlier, each packet copy may include information about the resources allocated for up to two subsequent replicas within its \ac{SCI}. Although increasing the number of replicas can improve reliability, it can also result in higher congestion and interference. Therefore, determining the optimal number of retransmissions requires balancing channel load and the desired reliability level. 

\subsection{Related work and contribution}

Mode~2, as a distributed scheduling algorithm, aims to minimize reciprocal user interference by having each user perform channel sensing and broadcast information about future resource reservations and occupations. Nevertheless, these techniques are not flawless, and packet collisions between users can still occur. An alternative approach is to cancel the resulting interference, which is the foundational principle of \ac{NOMA} and \ac{NOMA}-based methods. \ac{NOMA} has gained significant attention over the past decade as a potential solution to spectrum scarcity and the growing demand for higher throughput in cellular networks \cite{8357810,7676258}. However, as discussed and reviewed in \cite{10078378}, its application to the V2X sidelink has only recently been explored. Early works include \cite{di2017,9287485}, with initial investigations applying power-domain \ac{NOMA} to the 3GPP specifications detailed in \cite{hirai2019,todisco2023sic,rajalakshmi2024enhancing}.

NR-V2X sidelink Mode~2 differs from typical cellular networks,  as it operates in a dynamic, decentralized environment characterized by all-to-all communication. In our work, we do not enforce non-orthogonal transmissions, as is common in classical \ac{NOMA} implementations. Instead, we leverage the concept of \ac{SIC} from \ac{NOMA} to address interference. Specifically, we propose using \ac{SIC}-capable receivers to mitigate interference caused by overlapping transmissions in Mode~2 resource allocation. This interference becomes particularly critical under high data traffic conditions, with multiple transmissions and retransmissions, as expected in scenarios with a high penetration rate of \acp{CAV} running advanced applications.

Another source of inspiration for our work is the use of pointers to multiple packet repetitions in some \ac{NOMA}-based protocols, such as updated versions of the slotted ALOHA protocol. In 2007, Casini et al. proposed an enhancement to slotted ALOHA in \cite{casini2007}, introducing the transmission of multiple copies of the same packet, with each copy carrying a pointer to the slots containing the other copies. These pointers could then be used to identify and cancel interference caused by undetected copies, resulting in the concept known as \ac{CRDSA}. The true potential of CRDSA was revealed with the later work by Liva  \cite{liva2011}, which identified the procedure of \ac{SIC} in the context of random access with serial message-passing decoding of codes represented by sparse bipartite graphs over the binary erasure channel. The analogy of the approach with modern coding theory laid to the often used term coded-slotted ALOHA \cite{7302046}. 
The assumptions made in \cite{liva2011,7302046} and most of related work is that \textit{(i)} collisions are always destructive, i.e., a packet that interferes with another in a slot cannot be decoded; \textit{(ii)} the decoding is ideal in the absence of interference; and \textit{(iii)} \ac{SIC} is always ideal.

Recently, \ac{SIC} has primarily been applied to power-controlled transmissions in the uplink. However, the assumptions of a single receiver and uniform power levels do not hold in the V2X sidelink context, which operates under fundamentally different conditions. In \cite{ivanov2015,ivanov2017}, the authors propose a solution to address uncoordinated all-to-all broadcast random access for periodic messages exchanged between \acp{VUE}. In this approach, each \ac{VUE} generates multiple copies of its messages, randomly allocates them to different time slots, and broadcasts them to neighboring \acp{VUE}. However, the authors focus on a simplified scenario where nodes are uniformly distributed in a 2-D area, 
do not consider the impact of nodes distribution with realistic path-loss, 
and they assume a slotted timeline without considering the 3GPP sidelink specifications. 

Here, we propose and evaluate a novel solution that integrates 5G NR-V2X Mode~2 with the principles of \ac{CRDSA}. Through large-scale simulations in realistic highway scenarios, we demonstrate significant improvements in all-to-all broadcast transmissions, which are necessary for most V2X sidelink use cases. 

\section{RB-NOMA}\label{sec:receivers}

In this work, we propose VUEs to leverage advanced receivers capable of performing three operations: 1) cancellation of the signal component related to a decoded packet, which is \textit{\ac{SIC}}; 2) cancellation, in the current \ac{TTI}, of signals reconstructed from previously decoded copies of the same packet, called \textit{\ac{FRC}}; and 3) cancellation of signals reconstructed from copies in past TTIs, called \textit{\ac{BKC}}.
In the following, the three processes are detailed. A simplified block scheme of the corresponding receivers is reported in Fig.~\ref{fig:rx_schemes}. 

\subsection{Receiver with successive interference cancellation}\label{subsec:SIC}

\begin{figure}[t] 
	\centering
      \begin{subfigure}[]{0.48\textwidth}
         \centering
         \includegraphics[width=1\columnwidth]{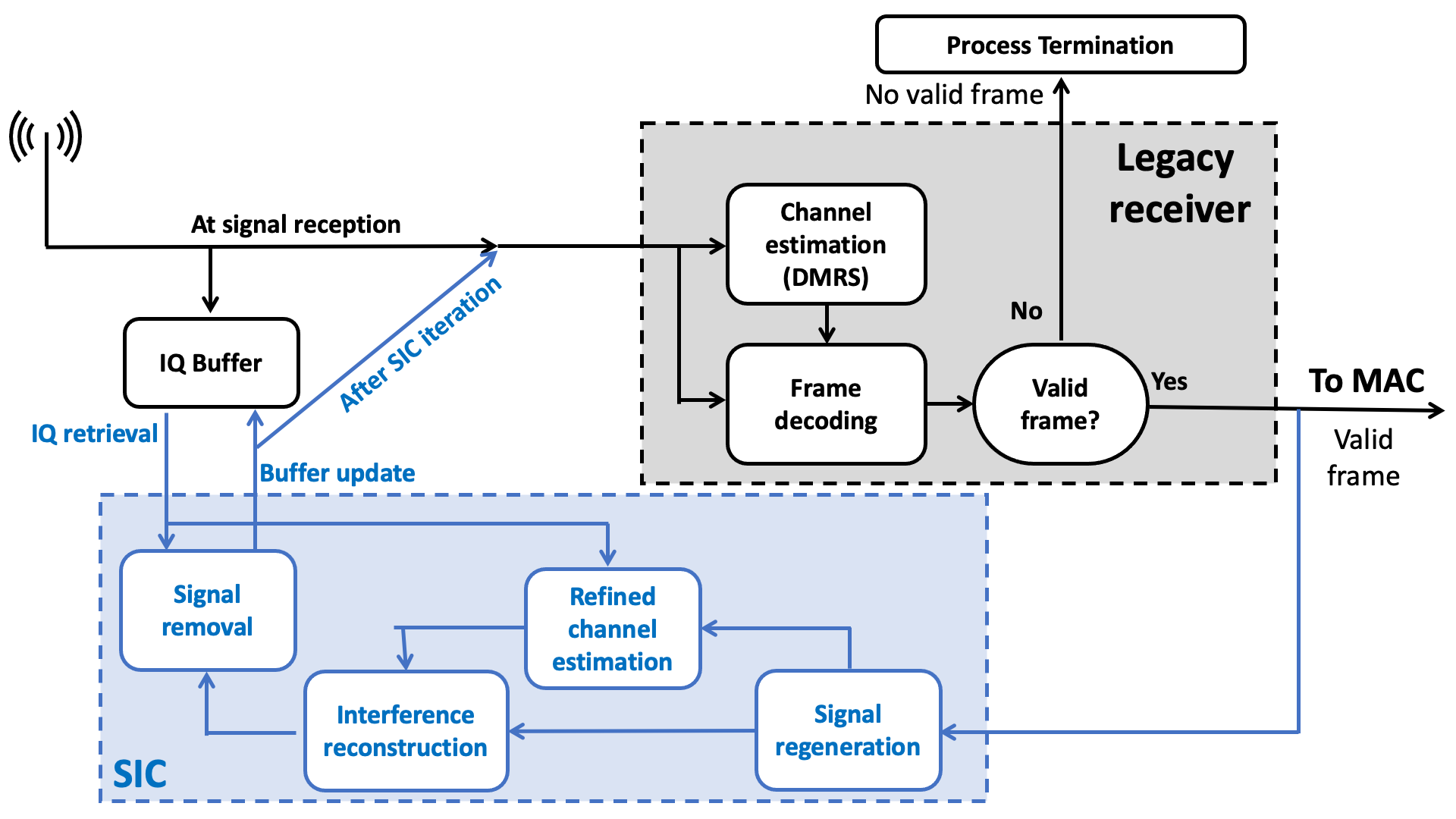} 
         \caption{SIC-aided receiver.}
          \label{fig:rx_schemes_sic}
     \end{subfigure}
     \vskip 0.4cm
      \begin{subfigure}[]{0.48\textwidth}
         \centering
         \includegraphics[width=1\columnwidth]{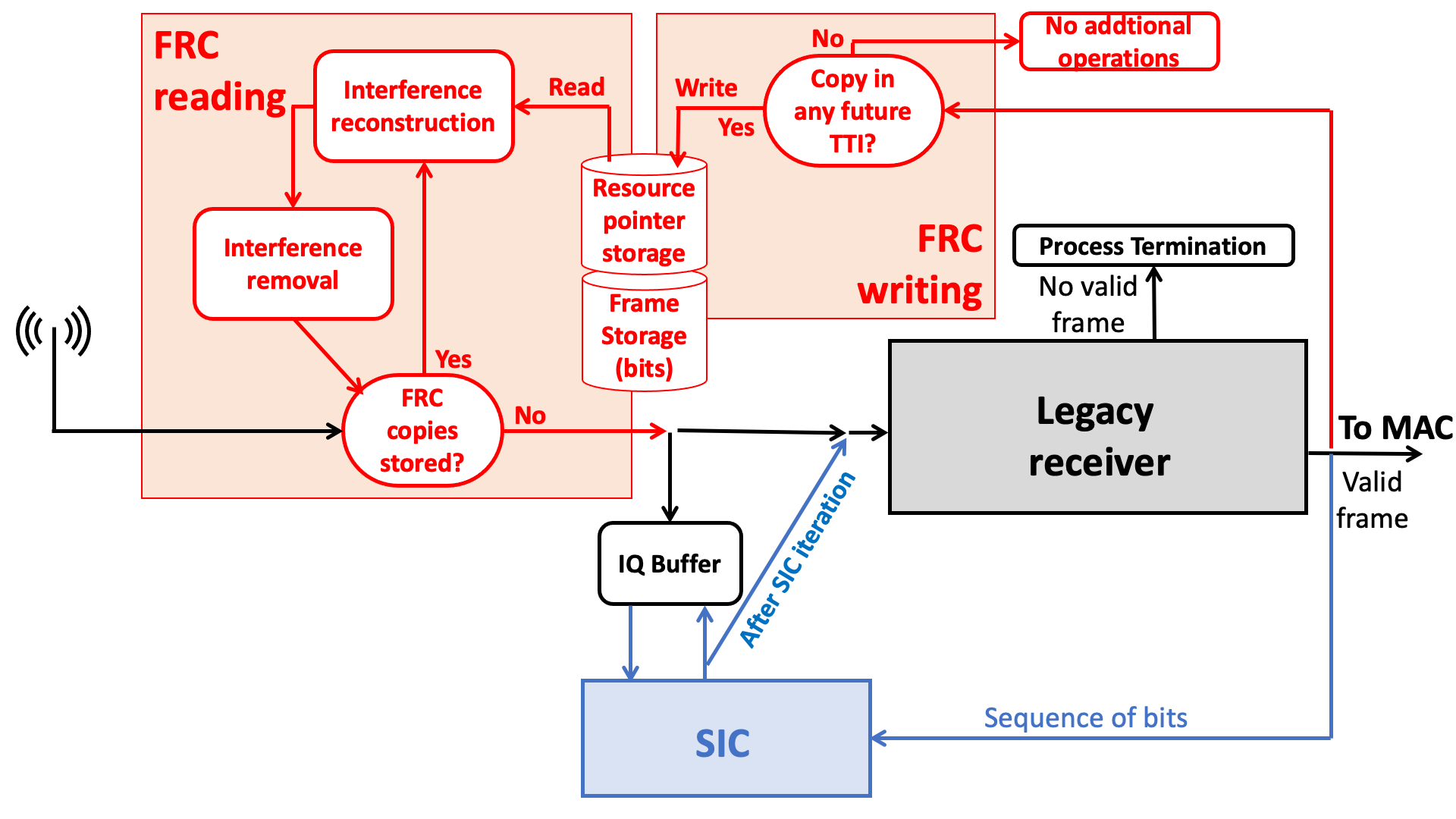} 
         \caption{SIC+FRC receiver.}
          \label{fig:rx_schemes_fwc}
     \end{subfigure} 
     \vskip 0.4cm
      \begin{subfigure}[]{0.45\textwidth}
         \centering
         \includegraphics[width=1\columnwidth]{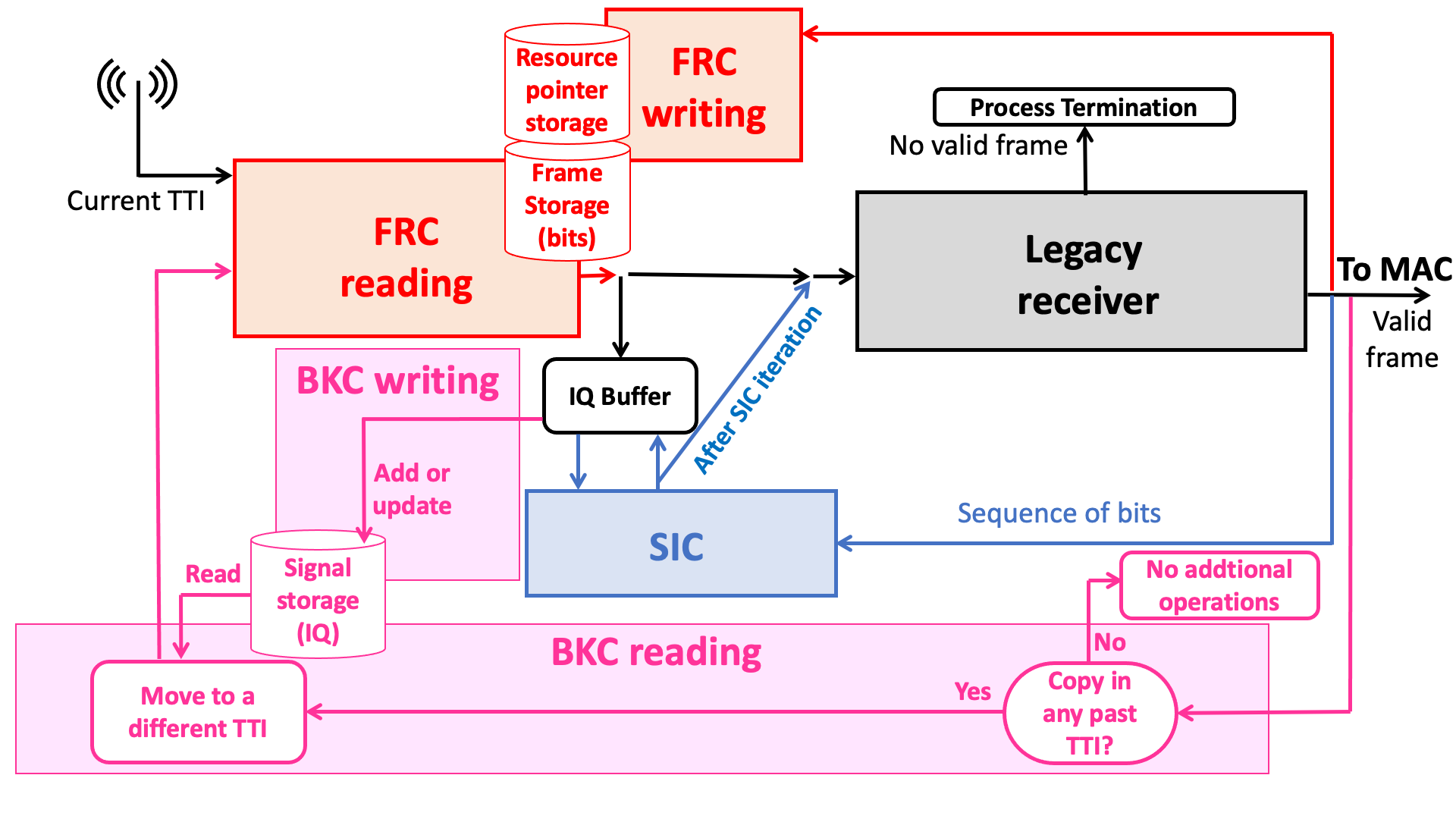}
         \caption{SIC+\ac{FRC}+\ac{BKC} receiver.}
          \label{fig:rx_schemes_bkc}
     \end{subfigure}  
     \caption{Block schemes of the compared receivers.}
	\label{fig:rx_schemes}
    \vskip -0.2cm
\end{figure}

Our solution builds on the \ac{SIC}-aided receiver, which is exemplified in Fig.~\ref{fig:rx_schemes_sic}. The upper part, inside grey dashed box, represents a legacy (i.e., non-SIC) receiver. The receiver first processes the received signal, using the \ac{DMRS} to estimate the channel and decode the packet. If a valid packet is not detected (due to the absence of a valid signal or a \ac{CRC} failure), the process terminates. If a valid packet is decoded, it is passed to the \ac{MAC} layer for further processing.

In the case of the SIC-aided receiver, when a valid packet is detected, it is also used for interference regeneration and cancellation, as illustrated in the blue box at the bottom of Fig.~\ref{fig:rx_schemes_sic}. Since the packet has been correctly decoded, the receiver knows all the transmitted bits and can treat the entire packet as if it was composed of pilot symbols. This enables the receiver to reconstruct the entire signal associated with the packet and refine the channel estimation. Once the signal corresponding to the decoded packet is obtained and the channel is estimated, it can be subtracted from the received signal previously stored in the IQ buffer. The result is a residual signal, which may contain another packet. The receiver then searches for this possible additional packet by reprocessing the residual signal through the grey box in Fig.~\ref{fig:rx_schemes_sic}. A mathematical description of the SIC process is detailed in Appendix~A.

\subsection{Receiver with forward cancellation}\label{WP2_solutions_FRC}

When multiple copies of the same packet are transmitted, the receiver can utilize one detected copy in a generic \ac{TTI} to estimate the signal associated with each copy in successive \acp{TTI} and remove it from the overall received signal in those \acp{TTI}. This type of receiver, depicted in Fig.~\ref{fig:rx_schemes_fwc}, also incorporates the functionalities of the \ac{SIC} block.

The forward cancellation process operates as follows:
\begin{itemize}
\item When the receiver decodes a packet, it checks the associated \ac{SCI} to determine if (and where) copies of that packet \textit{are planned in future \acp{TTI}} (FRC writing box in red in Fig.~\ref{fig:rx_schemes_fwc}).
\item If one or more copies are expected in future \acp{TTI}, the receiver stores the packet in a buffer labeled \textit{Frame Storage} within the FRC writing box in Fig.~\ref{fig:rx_schemes_fwc}. Only the bits of the packet need to be stored, not the signal itself. Additionally, the receiver stores pointers to the \acp{TTI} and subchannels where the replicas are expected, as indicated in the \ac{SCI} (see Section~\ref{sci-mod}), in the \textit{Resource Pointer Storage}.
\item In each subsequent \ac{TTI} and subchannel where a copy is expected (i.e., when copies actually arrive), the receiver uses the stored bits of the decoded packet to regenerate the transmitted signal. With the entire transmitted signal reconstructed, the channel can be estimated and fine-tuned, and the received signal can be recovered (FRC reading box in red in Fig.~\ref{fig:rx_schemes_fwc}).
\item Once the received signal corresponding to the copy is recovered, it is canceled from the overall received signal in the current \ac{TTI}. The residual signal then undergoes processing through the legacy receiver (gray box in Fig.~\ref{fig:rx_schemes_fwc}), potentially followed by the \ac{SIC} process (blue box in Fig.~\ref{fig:rx_schemes_fwc}) if a packet is decoded.
\end{itemize}
Please note that, since the resources used for the retransmissions are indicated in the SCI, if the SCI of none of the copies is correctly decoded, FRC cannot be applied.

\subsection{Receiver with backward cancellation}\label{WP2_solutions_BKC}

A receiver can also perform the cancellation operation backwards in time, enabling the recovery of additional packets. To achieve this, the receiver continuously stores the IQ samples in a buffer referred to as the \textit{Signal Storage} for a specified number of past \acp{TTI} and processes them through the blocks depicted in Fig.~\ref{fig:rx_schemes_bkc}. Given that the complexity of \ac{BKC} is higher than that of \ac{FRC}, we assume that \ac{FRC} is also implemented.

In more detail, the backward cancellation process operates as follows:
\begin{itemize}
    \item When the receiver decodes a packet, it checks the associated \ac{SCI} to determine if (and where) copies of that packet \textit{were transmitted in past \acp{TTI}}, as previously stored in the \textit{Signal Storage} (\ac{BKC} writing pink box in Fig.~\ref{fig:rx_schemes_bkc}).
    \item If one or more copies were transmitted in the past, the receiver proceeds to the \ac{BKC} reading pink box and uses the bits of the recently decoded packet to regenerate the entire signal for backward interference cancellation.
    \item The receiver then retrieves the recorded IQ samples from the \acp{TTI} associated with the transmitted copies.
    \item For each \ac{TTI} where a copy was sent, the receiver uses the regenerated signal to refine the channel estimation based on the recorded IQ samples. Since the transmitted signal corresponding to the copy is entirely known, the channel estimation can be fine-tuned, allowing the receiver to accurately estimate the received signal associated with the past copy.
    \item Once the received signal corresponding to the copy is estimated, it is canceled from the overall IQ samples. The residual signal is then reprocessed to detect and decode any additional packets that may have been transmitted in the same \ac{TTI}.
\end{itemize}
Also in this case, if none of the SCI of the copies is correctly decoded, the resources used for previous or later repetitions is unknown, and neither FRC nor BKC can be applied. 
A possible implementation of the operations of the RB-NOMA receiver with SIC+FRC+BKC is detailed in Algorithm~\ref{alg:rb-noma}. 

\begin{algorithm}[t]
\footnotesize
\caption{RB-NOMA with FRC and BKC}
\label{alg:rb-noma}
\begin{algorithmic}[1]
\State  \textbf{Notation $\mathcal{F}(t)$:} Element of the frame storage at TTI $t$ 
\State  \textbf{Notation $\mathcal{Q}(t)$:} Element of the signal storage at TTI $t$  
\State  \textbf{Variables:} Current TTI $t_0$, elaborated TTI $t$, current signal $s_{t_0}$, elaborated \hbox{signal $s$}
\State  \textbf{Step 1:} Initialize $t \gets t_0$, $s \gets s_{t_0}$
    \State  \textbf{Step 2:} $\forall$ packet $p \in \mathcal{F}(t)$: Reconstruct and cancel the signal corresponding to $p$ from $s$
    \State  \textbf{Step 3:} $s \to \mathcal{Q}(t)$
    \State  \textbf{Step 4:} Attempt decoding a new packet from $s$
    \If{a new packet $p$ is decoded}
        \State  \textbf{Step 5:} Read the SCI of $p$
    \For{each pointer to a repetition $r$}
        \If{$r$ is in $t_n > t_0$ (future)} 
            \State  store $r \to \mathcal{F}(t_n)$ 
        \Else 
            \State  $s' \gets \mathcal{Q}({t_n})$
            \State  Reconstruct and cancel the signal corresponding to $r$ from $s'$
            \State  $s' \to \mathcal{Q}({t_n})$
            \State  Start a parallel process from Step 2 with $t \gets t_n$, $s \gets s_{t_n}$
        \EndIf
    \EndFor
        \State  Reconstruct and cancel the signal corresponding to $p$ from $s$
        \State  Return to Step 3
    \Else
        \State  \textbf{End process}
    \EndIf
\end{algorithmic}
\end{algorithm}

\subsection{A toy example}

\begin{figure}[t]
\centering
\includegraphics[width=1\columnwidth]{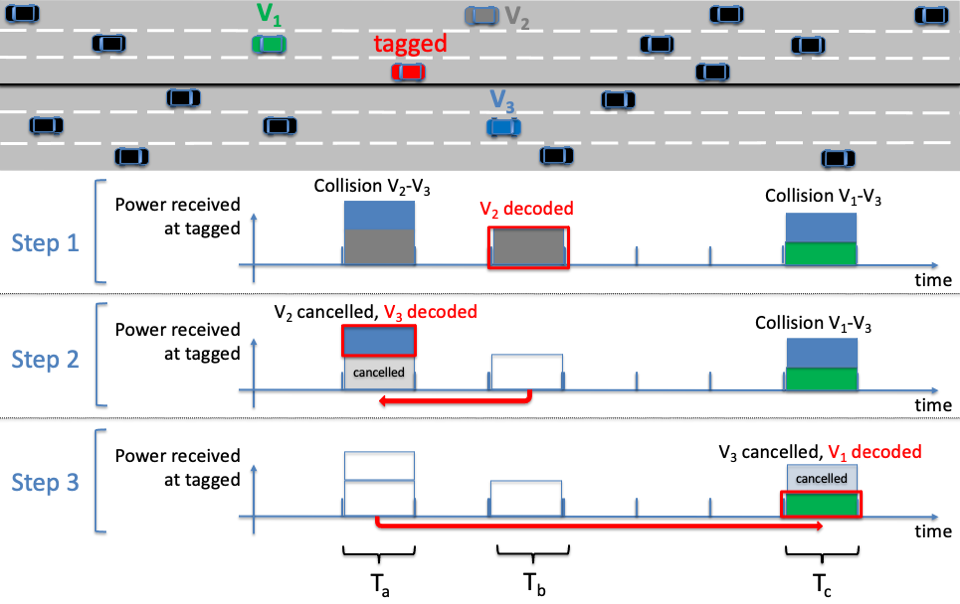}
\caption{Example of joint FRC and BKC process. Packets (and relevant copies) transmitted by V$_1$, V$_2$, V$_3$ are reported in green, gray, and blue, respectively.}
\label{fig:ExampleBKC}
\end{figure}

An example of the behavior of a receiver with \ac{FRC} and \ac{BKC} is illustrated in Fig.~\ref{fig:ExampleBKC}, where a \ac{VUE} receives simultaneous transmissions from neighboring \acp{VUE}.
In the figure, the tagged vehicle (red) receives packets broadcast by V$_1$ (green), V$_2$ (gray), and V$_3$ (blue) over three TTIs, labeled T$_\text{a}$, T$_\text{b}$, and T$_\text{c}$. Mathematically, the samples of the signal received by the tagged vehicle in these \acp{TTI} can be written, respectively, as: 
\begin{align}\label{eq:signalsTagged_Tabc}
    y(\text{T}_\text{a},k) &= h_{\text{V}_2}(\text{T}_\text{a},k) x_{\text{V}_2}(\text{T}_\text{a},k) \nonumber\\&+ h_{\text{V}_3}(\text{T}_\text{a},k) x_{\text{V}_3}(\text{T}_\text{a},k) +  n(\text{T}_\text{a},k) \\
      y(\text{T}_\text{b},k) &= h_{\text{V}_2}(\text{T}_\text{b},k) x_{\text{V}_2}(\text{T}_\text{b},k) +  n(\text{T}_\text{b},k) \\    y(\text{T}_\text{c},k) &= h_{\text{V}_1}(\text{T}_\text{c},k) x_{\text{V}_1}(\text{T}_\text{c},k) \nonumber\\&+ h_{\text{V}_3}(\text{T}_\text{c},k) x_{\text{V}_3}(\text{T}_\text{c},k) +  n(\text{T}_\text{c},k)  
\end{align}
where $k$ is used to indicate the $k$-th sample within the given TTI, $x_{\text{V}_j}$ is the signal received from the VUE $\text{V}_j$, $h_{\text{V}_j}$ is the channel coefficient from the VUE $\text{V}_j$ to the tagged VUE, and $n$ is the \ac{AWGN} contribution.

During the TTI T$_\text{a}$, the receiver is unable to decode either of the two signals because the interference they reciprocally cause is too large. The station stores, in the Signal Storage, the samples $y(\text{T}_\text{a},k)$ associated with the TTI T$_\text{a}$.

Then, during the TTI T$_\text{b}$, the receiver decodes the packet sent from V$_2$, which includes the SCI and the data. From the SCI, the receiver knows that a copy of the same packet had been received during the TTI T$_\text{a}$ and retrieves the signal $y(\text{T}_\text{a},k)$ from the Signal Storage. Using the decoded packet and the information in the SCI, the receiver perfectly reconstructs $x_{\text{V}_2}(\text{T}_\text{a},k)$ and uses it with $y(\text{T}_\text{a},k)$ to estimate $h_{\text{V}_2}(\text{T}_\text{a},k)$. Please notice that, even if the design of the specific channel estimation algorithm is beyond the scope of this work, given that $x_{\text{V}_2}(\text{T}_\text{a},k)$ is known, the estimation is performed as there were pilot symbols sent over all the subcarriers and all the OFDM symbols of the TTI, which means that the estimation can be accurate (the inaccuracy of the channel estimation is anyway taken into account in the simulations). 
Denoting the estimation as $\hat{h}_{\text{V}_2}(\text{T}_\text{a},k)$, the receiver performs the interference cancellation of the signal from V$_2$ and obtains: 
\begin{align}\label{eq:signalsTagged_TaRes}
    y(\text{T}_\text{a},k) &= h_{\text{V}_3}(\text{T}_\text{a},k) x_{\text{V}_3}(\text{T}_\text{a},k) \\ &+ \left(h_{\text{V}_2}(\text{T}_\text{a},k)-\hat{h}_{\text{V}_2}(\text{T}_\text{a},k) \right)x_{\text{V}_2}(\text{T}_\text{a},k) +   n(\text{T}_\text{a},k) \;.  \nonumber
\end{align}
It is assumed that the residual interference from V$_2$ is now sufficiently low to allow the receiver of the tagged vehicle to also decode the packet from V$_3$. Decoding the packet and reading the SCI, the receiver also knows that a copy of that packet will be received in the TTI T$_\text{c}$, and therefore stores the packet and the reference to that TTI in the Frame Storage.

During the TTI T$_\text{c}$, based on the recorded information, the receiver  
retrieves the packet received from V$_3$, reconstructs $x_{\text{V}_3}(\text{T}_\text{c},k)$, estimates $\hat{h}_{\text{V}_3}(\text{T}_\text{c},k)$, and obtains: 
\begin{align}\label{eq:signalsTagged_TcRes}
    y(\text{T}_\text{c},k) &= h_{\text{V}_1}(\text{T}_\text{c},k) x_{\text{V}_1}(\text{T}_\text{c},k) \\ &+ \left(h_{\text{V}_3}(\text{T}_\text{c},k)-\hat{h}_{\text{V}_3}(\text{T}_\text{c},k) \right)x_{\text{V}_3}(\text{T}_\text{c},k) +   n(\text{T}_\text{c},k) \;.  \nonumber
\end{align}
The residual interference from V$_3$ is now   sufficiently low to allow the receiver of the tagged vehicle to also decode the packet from V$_1$.

In this example, the tagged VUE is able to decode all three packets thanks to FRC and BKC, whereas a legacy receiver could only decode the packet from V$_2$.

\subsection{Complexity, SCI modifications, and memory requirements}\label{sci-mod}

\textbf{Complexity.} The complexity of our proposal mainly depends on the SIC-based decoding of overlapping signals at the receiver. The reading and writing operations from buffers, which are added by FRC and BKC, are in fact of negligible impact (also considering that the buffer size is limited by the delay budget, normally bounded to 100~ms or less). Detailed considerations about the computation complexity of SIC require the definition and/or identification of specific hardware and software implementations, which are beyond the scope of our study. The proposal appears feasible considering the size and power availability of vehicles, and given the modern on-board computing platforms.

In any case, one design choice that can help reduce the complexity is to limit the number of iterations. These, which mean reconstructions and cancellations, are in fact the main parameter affecting the processing burden  \cite{almohamad2020low} and the decoding time \cite{manglayev2017noma}, \cite{saito2013non}. In particular, in this work it is assumed that at most a single iteration is performed; this choice follows \cite{todisco2023sic}, where results show that a single iteration already brings substantial performance gains in vehicular scenarios.

\textbf{Overhead.} With the proposed forward and backward cancellation techniques, the receiver needs to know where packet copies will be and have been transmitted, respectively. This information is written in the SCI of the other copies and to this aim we extend the SCI to carry additional bits.

In particular, for the forward cancellation the \ac{SCI} associated with each packet needs to include an indication of the subchannels and \acp{TTI} where the future packet copies will be sent. This is foreseen in \textit{legacy NR only for at most the first two upcoming copies} \cite{38.212}.  Hence, the \ac{SCI} should be  modified to additionally carry the time and frequency position of all future remaining packet repetitions after the first two. Denoting as $N_\text{copies}$ the number of copies, this implies no more than $(N_\text{copies}-2) \log_2 ( {N_\text{dTTI}}_\text{max} \cdot N_\text{subch} )$ additional bits, where ${N_\text{dTTI}}_\text{max}$ is the maximum number of TTIs between the first transmission and the last copy, and $N_\text{subch}$ is the number of subchannels. If we assume, as an example, that $N_\text{copies}=3$, ${N_\text{dTTI}}_\text{max}=32$, and $N_\text{subch}=10$ (like in the simulations hereafter), we need 9 additional bits, which appears negligible. 

To implement backward cancellation, the \ac{SCI} associated with each packet needs to include an indication of the subchannels and \acp{TTI} where the past copies were sent, which is not foreseen in 3GPP specifications. This additionally entails the transmitter to keep track of such information per each packet. If FRC and BKC are implemented, all packets need to carry information about all the other copies (earlier or later). Adding one bit per each repetition to indicate if it is backward or forward, the resulting additional number of bits is equal to $(N_\text{copies}-2) \log_2 ( {N_\text{dTTI}}_\text{max} \cdot N_\text{subch} ) + N_\text{copies}$. Assuming at most 3 retransmissions as in Section \ref{sec:results}, the same assumptions as in the FRC case bring to 9 bits required for the additional pointer (for the third retransmission) and 3 bits to indicate the direction (per each of the three retransmissions), which means 12 additional bits. 

\textbf{Memory.} RB-NOMA has also limited requirements in terms of memory, especially when looking at SIC and FRC. Looking at the buffer required for SIC, if we assume IQ samples of 4~bytes (16 bits per via) and we consider the Nyquist sampling rate of 40 Ms/s for a 20 MHz bandwidth, we obtain for 1 ms TTI (i.e., $\mu$=0) a requirement of 160 kbyte. Regarding the size of the frame storage for FRC, if we assume to record 50 packets of 1000 bytes each, we need additional 50 kbyte and a few bits per packet for the resource pointer storage, for an overall requirement close to 210 kbyte. Finally, adding the signal storage of BKC, if we assume IQ samples for 32 TTIs we obtain an additional 5.12 Mbyte, which brings the total requirement to approximately 5.4 Mbyte. The SCI overhead and memory requirements are summarized in Table~\ref{Tab:overheadRev}.

\begin{table*}[t]
\caption{Additional SCI bits of overhead  and memory requirements.}
\vspace{-2mm}
\label{Tab:overheadRev}
\small
\centering
\begin{tabular}{p{2.7cm}p{5.4cm}p{3.7cm}p{2cm}}
\hline \hline\textbf{Scheme} & \textbf{Max. additional SCI overhead (generic)}& \textbf{Additional SCI overhead (our settings)} & \textbf{Buffer size} \\ \hline
 SIC & - & - &  160 kbytes  \\
SIC+FRC &  $(N_\text{copies}-2) \log_2 ( {N_\text{dTTI}}_\text{max} \cdot N_\text{subch} )$ & 9 bits & 210 kbytes \\
SIC+FRC+BKC &  $(N_\text{copies}-2) \log_2 ( {N_\text{dTTI}}_\text{max} \cdot N_\text{subch} ) + N_\text{copies}$ & 12 bits & 5.4 Mbytes\\
\hline
\end{tabular}
\end{table*}

\section{Abstraction of the physical layer}\label{sec:SICmodeling}

In this paper, we aim to investigate the network-level performance of RB-NOMA through large-scale simulations with hundreds of vehicles. This requires an abstraction of the physical layer, able to model the presence of interference and the impact of cancellation.

Let us focus on a generic VUE, let's say $i$, receiving in a given TTI. The neighboring \acp{VUE} transmitting in the same TTI are $\text{u}_1$, $\text{u}_2$, ..., $\text{u}_{|\Uset_i|}$, where $\Uset_i$ is the set of all \acp{VUE} transmitting in the neighborhood of vehicle $i$ in the considered time slot and $|\Uset_i|$ is the cardinality of the set $\Uset_i$.\footnote{The set $\Uset_i$ includes all \acp{VUE} that interfere significantly on $i$. Although the size of the set depends on the channel conditions and it is not known at the receiver, this is irrelevant for the SIC processing. The receiver in fact decodes and cancels one signal at a time irrespective of the number of remaining interfering signals.} Without loss of generality, let us assume that they are sorted so that, if $m<n$, the signal received from neighbor $\text{u}_m$ is stronger than the one received from user $\text{u}_n$.

\textbf{Legacy receiver.} Without SIC capability, the legacy receiver can decode only the packet received from the strongest source and thus, the only relevant source is $\text{u}_1$. The \ac{SINR} of the signal received from user $\text{u}_1$ can be written as 
\begin{equation}\label{NOMAgenericSINR1}
\gamma_{i,\text{u}_1}=\frac{\Prcv{i,\text{u}_1}}{\Pnoise+\sum_{j=2}^{|\Uset_i|}\IBEcoeff_{i,\text{u}_j}\Prcv{i,\text{u}_j}}\;,  
\end{equation}
where $\Prcv{i,\text{u}_j}$ is the power received at $i$ from $\text{u}_j$, $\Pnoise$ is the noise power and $\IBEcoeff_{i,\text{u}_j}$ is a multiplying coefficient, between 0 and 1, that quantifies how  much  power  is  sent  by $\text{u}_j$ in  the  bandwidth  used  by $i$, related to the power transmitted by $\text{u}_j$; $\IBEcoeff_{i,\text{u}_j}$ is 1 if $\text{u}_j$ uses exactly  the  same  bandwidth as $i$,  and  is  lower  than~1, also depending on the modeling of the \ac{IBE} \cite{38.101-1}, if its signal does not overlap or it overlaps only partially. With the settings assumed in this work, the signals use the full bandwidth and, therefore, $\IBEcoeff_{i,\text{u}_j}$ is always equal to~1.

\textbf{Receiver with SIC only.} When SIC is implemented, if the packet sent by the strongest source ($\text{u}_1$) is decoded, then the cancellation of the corresponding signal is performed and the decoding process is attempted again to possibly decode the packet sent by the second strongest source ($\text{u}_2$), and so on. Formally, we can write that at user $i$ the \ac{SINR} related to the signal received from user $\text{u}_z \in \Uset_i$ can be written as \begin{equation}\label{NOMAgenericSINR2}
    \gamma_{i,\text{u}_z}=\frac{\Prcv{i,\text{u}_z}}{\Pnoise+\kNOMA \sum_{j=1}^{z-1} \IBEcoeff_{i,\text{u}_j}\Prcv{i,\text{u}_j}+\sum_{j=z+1}^{|\Uset_i|}\IBEcoeff_{i,\text{u}_j}\Prcv{i,\text{u}_j}}  \;,
\end{equation}
where $\kNOMA\in[0,1]$ determines the interference cancellation capability ($\kNOMA=0$ for ideal SIC and $\kNOMA=1$ for no SIC). In the denominator of \eqref{NOMAgenericSINR2}:
\begin{itemize}
    \item The first term is the noise power;
    \item The second term is the residual interference from the stronger interfering signals, which are possibly decoded before this one;
    \item The third term is the interference from the weaker interfering signals, which cannot be canceled.
\end{itemize}

In \eqref{NOMAgenericSINR2}, we consider through the parameter $\kNOMA$ non-ideal SIC where only a portion of the user's signal is removed, leaving residual interference. Common causes of residual interference are imperfect users' channel estimation and approximate estimation of user modulation symbols \cite{novak2013idma}. We model residual interference as additive and Gaussian, with zero mean and variance determined by the received power minus the SIC capability expressed in dB \cite{tweed2018dynamic}.

The receiver processes the signals from the strongest down to the weakest and the process ends as soon as a signal is not decoded. Assuming that the signal is correctly decoded if the \ac{SINR} is above a given threshold $\SINRt$ \cite{WuBarMarBaz:J22}, the correctness of the reception at the generic user $i$ from the generic user $u_z \in \Uset_i$ can be indicated as a Bernoulli variable equal to
\begin{equation}
    \delta_{i,u_z} = \begin{cases}
1 & \text{if~$\delta_{i,u_{z-1}}=1$~AND~$\gamma_{i,u_z} \geq \SINRt$}\\
0 & \text{otherwise}\\
\end{cases}~\forall z\in\{1,|\Uset_i|\}\;,
\end{equation}
where $\delta_{i,u_{0}}$ is conventionally equal to 1.

\textbf{Receiver with RB-NOMA.} 
When \ac{FRC} or \ac{BKC} are also implemented in addition to SIC, it may be that the interference from a source that is not stronger than the one under elaboration has been already cancelled. 
Thus, 
the \ac{SINR} related to the signal received from user $\text{u}_z \in \Uset_i$ 
needs to be rewritten in a more general way, as 
\begin{equation}\label{NOMAgenericSINR3}
\gamma_{i,\text{u}_z}=\frac{\Prcv{i,\text{u}_z}}{\Pnoise+\sum_{j=1,j\neq z}^{|\Uset_i|} ((1-\xi_{i,\text{u}_j}) + \kNOMA \xi_{i,\text{u}_j}) \IBEcoeff_{i,\text{u}_j}\Prcv{i,\text{u}_j}}\;,   
\end{equation}
where $\xi_{i,\text{u}_j}$  is a Bernoulli binary variable, called cancellation variable, that indicates if the signal received from user $\text{u}_j$ has been cancelled or not. If the signal has been cancelled, then $\xi_{i,\text{u}_j}=1$ and the power received from $\text{u}_j$ has been (possibly partially) removed from the denominator; otherwise, $\xi_{i,\text{u}_j}=0$ and the entire corresponding power contribution remains.

\section{Performance evaluation}\label{sec:results}

The performance of the proposed RB-NOMA is hereafter assessed using the open source WiLabV2Xsim\footnote{The simulator is available at \url{https://github.com/V2Xgithub/WiLabV2Xsim}.} \cite{todisco2021performance}.
The simulator, developed in MATLAB  to study the delivery performance of data exchanged over NR-V2X sidelink, has been overhauled to model the advanced receivers with \ac{SIC}, \ac{FRC}, and \ac{BKC}, following the details provided in Section~\ref{sec:receivers}. The main simulation settings are summarized in Table~\ref{Tab:Settings_MS} and further discussed in the following. 
The implementation of the RB-NOMA reception process in the simulator is described in Appendix~B. 

\subsection{Simulation settings}\label{sec:WP22_simsettings}

\begin{table}[t]
\caption{Main simulation settings. (*) used when not differently specified. }
\vspace{-2mm}
\label{Tab:Settings_MS}
\footnotesize
\centering
\begin{tabular}{p{4.2cm}p{3.4cm}}
\hline \hline
\emph{\textbf{Road traffic}} & \\
Road layout & Highway, 3+3 lanes \\
Density & Variable \\
Average speed & 70 km/h \\ \hline
\emph{\textbf{Data traffic generation}}& \\
Inter-packet generation, periodic & 100 ms \\ 
Inter-packet generation, aperiodic & 50 ms + average 50 ms \\ 
Packet size & 1000~bytes \\ \hline
\emph{\textbf{Access layer}} & \\
Keep probability & 0 \\
RSRP sensing threshold of SB-SPS & -$126$~dBm  \\
Min. time for the allocation, $T_1$ & $1$~ms \\
Max. time for the allocation, $T_2$ & $50$~ms\\
Blind retransmissions & From 0 to 3 \\
\hline 
\emph{\textbf{Signal}} & \\
Channels & ITS bands at 5.9 GHz \\
Bandwidth & 20 MHz \\
SCS & 15 kHz (*) \\
Subchannel size & \mbox{10 PRB} \\
Number of subchannels & 10 \\
MCS & 5 (4-QAM, $R_\text{c}=0.37$) (*)\\
SINR threshold &  3.6 dB (*) \\
\hline 
\emph{\textbf{SIC}} & \\SIC cancellation factor & -30 dB \\
Maximum number of SIC iterations & 1 \\
\hline
\emph{\textbf{Propagation}} & \\
Transmission power & 23~dBm \\
Antenna gain (tx and rx)  & 3 dBi \\
Noise figure & 9~dB \\
Propagation model & Modified ECC Rural \\
Shadowing & Var. 3 dB, decorr. dist. 25~m \\ \hline
\hline
\end{tabular}
\end{table}

\textbf{Road traffic.} As a reference scenario, we consider a highway with variable vehicle density, and with vehicles distributed across three lanes per direction. 
Each vehicle moves at a random speed defined according to a Gaussian distribution with a 70 km/h average and a 7 km/h standard deviation.\footnote{Additional simulations, not shown here for the sake of conciseness, observed that the results remain similar when the average speed is changed, provided the other settings are maintained.}

\textbf{Data traffic generation.} Each \ac{VUE} generates 1000 bytes-long packets to resemble the worst-case data traffic pattern identified for advanced \acp{CAV} applications, such as those focusing on the collective perception service in \cite{c2c-2020_spectrum}. Two types of traffic are considered:
\begin{itemize}
    \item \textit{Periodic traffic}: packets are generated every 100~ms; 
    \item \textit{Aperiodic traffic}: the time interval between the generation of one packet and the following one is a random variable consisting of the sum of a fixed part of 50~ms and a variable part with negative exponential distribution and average 50~ms; this generation follows 3GPP recommendations in \cite{3GPP37885}. 
\end{itemize}

We assume that the periodic traffic is served by SB-SPS and the aperiodic traffic by SB-DS.

\textbf{Access layer.}  
The keep probability $\pkeep$ is set to~0, meaning that a resource reselection is always performed when the reselection counter reaches zero, which occurs after an interval that varies randomly between 0.5 and 1.5~s. 
Subchannels are identified as busy when the sensed power is above an \ac{RSRP} threshold equal to -126~dBm, as justified in \cite{todisco2021performance}. 
The delay budget is constrained to 50~ms, which corresponds to the minimum inter-generation time in the aperiodic case. This ensures that each packet is sent before the next one is generated. The correct reception of each packet is detected based on a SINR threshold set to 3.6~dB \cite{WuBarMarBaz:J22}. 
For both periodic and aperiodic data, we assume that each vehicle repeats a packet transmission for a given number of blind retransmissions (up to 3), which is fixed and uniform across all vehicles.\footnote{Through additional simulations, we observed that assuming more than~3 retransmissions did not give further relevant improvements in any of the considered scenarios.}

\textbf{Signal.} Transmissions occur within the so-called 5.9$\;$GHz \ac{ITS} band in a 20~MHz channel bandwidth, which is currently the largest bandwidth being considered for sidelink communication on a global scale \cite{fcc}, with ten subchannels of 10 \acp{PRB} each. Numerology with parameter $\mu$ equal to 0 is assumed, implying SCS of $15\;$kHz and TTI duration of 1~ms. The \ac{MCS}~5 is used, corresponding to the most reliable one to allocate 1000~bytes in one TTI; this implies, under the considered settings, that each 1000 bytes-long packet transmission is performed over ten subchannels \cite{38.214}. 

\textbf{SIC.} A single SIC iteration is assumed, which means that only the strongest signal can be canceled, and the process is not repeated if another frame can be decoded after that cancellation; this choice is based on the results shown in \cite{todisco2023sic}, and avoids additional delay and complexity increase, being both dependent on the number of iterations \cite{manglayev2017noma}, \cite{saito2013non}. The interference cancellation capability $\kNOMA$ is set to -30~dB, based on the results shown in \cite{todisco2023sic}. 

\textbf{Propagation.} 
Consistently with real implementations, the transmission power is set to $23~$dBm, and the antenna gain is equal to $3~$dBi at both the transmitter and receiver sides. The noise figure of the receiver is assumed to be 9~dB. The path-loss model follows the modified ECC Report 68 rural described in \cite{etsi202110} and recently used in ETSI for similar studies, with correlated log-normally distributed shadowing, characterized by a standard deviation of 3~dB and a decorrelation distance of 25~m, as per 3GPP guidelines in \cite{3GPP_TR_36_885}. 

\begin{figure}[t]
\centering
\includegraphics[width=1\columnwidth]{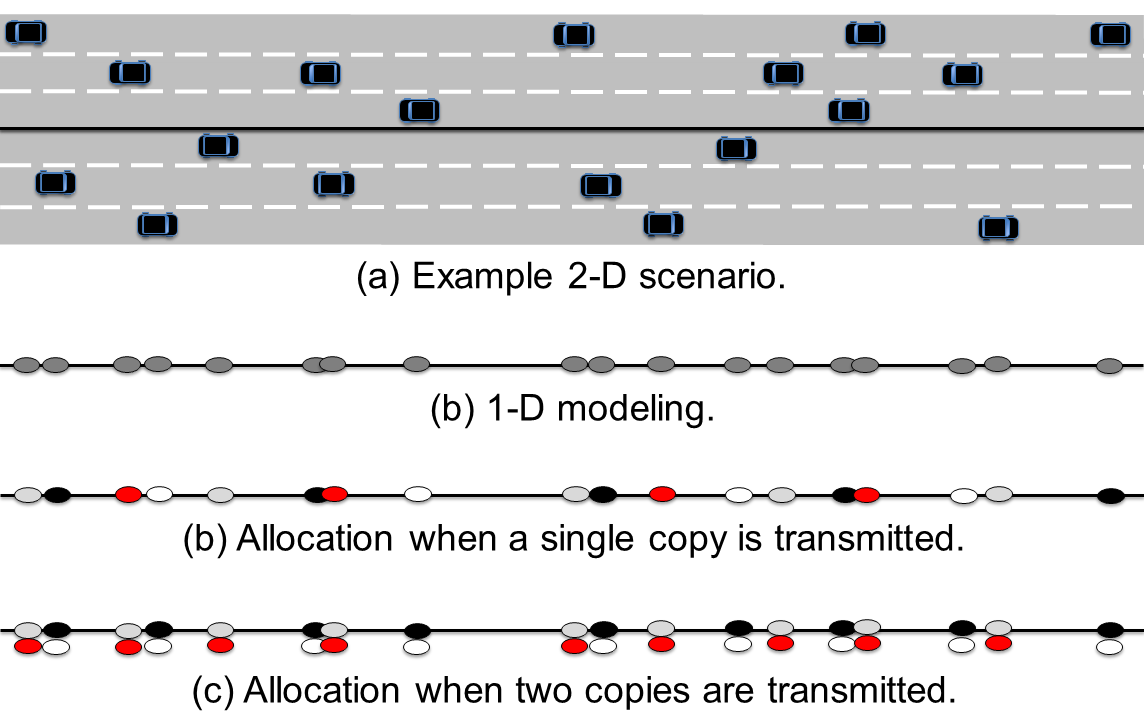}
\caption{Exemplification of the sorted allocation, assuming four orthogonal resources (gray, black, red, white).}
\label{fig:ExampleSorted}
\end{figure}

\subsection{Output metrics}
Performance is assessed in terms of the following metrics:
\begin{itemize}
    \item The \textit{\ac{CBR}}, calculated during observation windows of 100~ms 
as $\Nbusy/\Ncbr$, 
with $\Nbusy$ representing the number of subchannels where the sensed power is above a threshold of \hbox{-94~dBm} and $\Ncbr$  the total number of subchannels \cite{ETSI_103_574};
\item The \textit{\ac{PRR}}, computed as the ratio between the number of \acp{VUE} successfully receiving  the packets at a given distance from the transmitter and the total number of \acp{VUE} located at the same distance \cite{3GPP37885};
        \item The \textit{range}, which assesses the maximum distance over which messages are received with \ac{PRR}$>$0.95 as used by 3GPP in \cite{3GPP_TR_22_885}; 
\item The \textit{\ac{WBSP}}, defined as the probability  that, in an observation window, two vehicles cannot receive updates from each other (i.e., no messages are correctly received) \cite{bazzi2020wireless}. If several consecutive messages  are  lost  from  a  generic  vehicle,  the shared information is no longer updated for some time, and that time is here called \ac{WBS}. 
The \ac{WBSP} is calculated by considering messages exchanged among vehicles at reciprocal distances of no more 
than 100~m, as the exchange of messages with nearby vehicles is particularly crucial for safety purposes;
\item The \ac{EED}, defined, similarly to \cite{daw2024lamp}, as the delay the message incurs since it is generated at the application layer of source VUE up to the point the application layer of receiving VUE correctly receives it. When multiple copies of a message are sent to increase reliability, the reception of the first packet is considered in calculating the metric. The latency incurred in SIC operations at the receiver is not considered, since expected in the order of microseconds \cite{xu2021low}, \cite{sunkaraboina2024fpga}, and hence negligible compared to other contributions.
\end{itemize}

\begin{figure}[t]
\centering
\includegraphics[width=0.85\columnwidth]{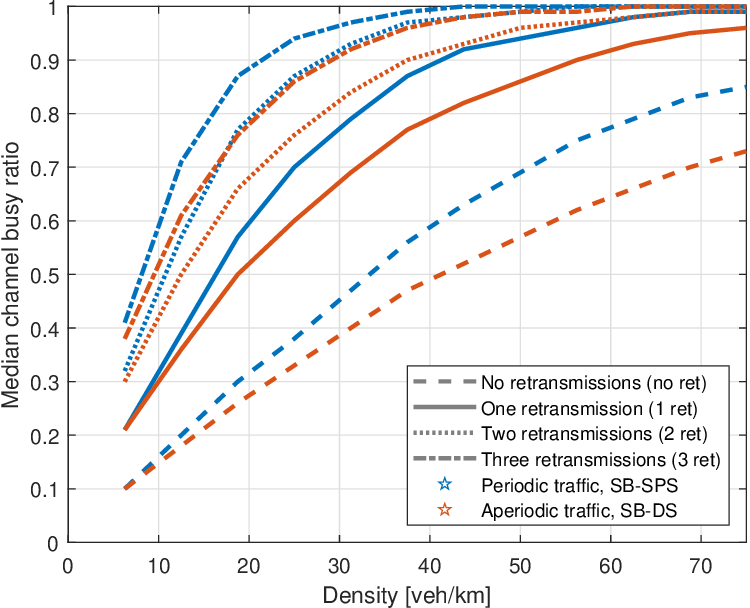}
\caption{Median channel busy ratio vs. vehicle density.}
\label{fig:CBR_MS}
\end{figure}

\subsection{Benchmarking schemes}

Within the simulation campaign, the following \textit{solutions} are compared:
\begin{itemize}
    \item Mode~2 with legacy receivers that do not implement SIC (or simply \textit{Legacy});
    \item Mode~2 with receivers that implement only SIC (shortened as \textit{SIC}), as also considered in \cite{10078378,todisco2023sic}; \item Enhanced Mode~2 with receivers that implement RB-NOMA with SIC and FRC (shortened as \textit{SIC+\ac{FRC}});
    \item Enhanced Mode~2 with receivers that implement RB-NOMA with SIC, \ac{FRC}, and \ac{BKC} (shortened as \textit{SIC+\ac{FRC}+\ac{BKC}}).    
\end{itemize}

\begin{figure*}[t] 
	\centering
      \begin{subfigure}[]{0.45\textwidth}
         \centering
         \includegraphics[width=0.9\columnwidth]{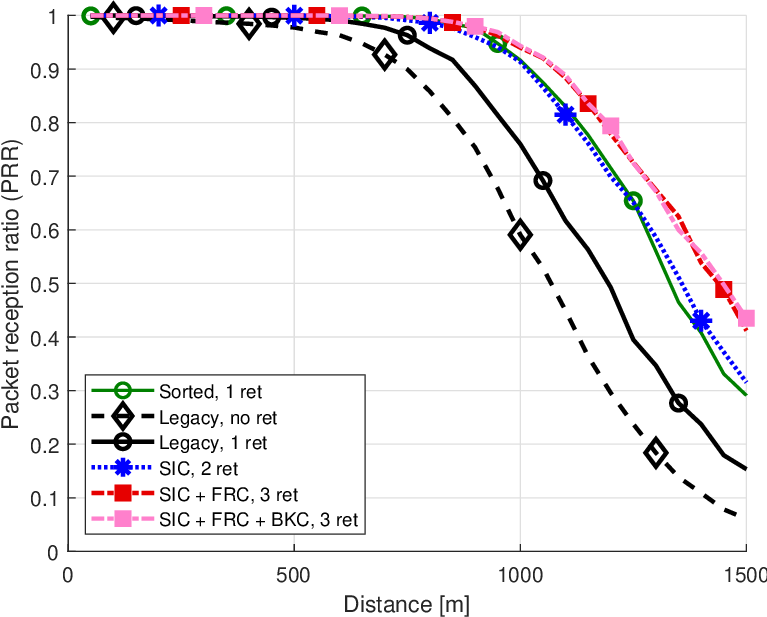}
         \caption{Periodic traffic, SB-SPS.}
          \label{fig:MS_PRR_50_red_a}
     \end{subfigure} 
      \begin{subfigure}[]{0.45\textwidth}
         \centering
         \includegraphics[width=0.9\columnwidth]{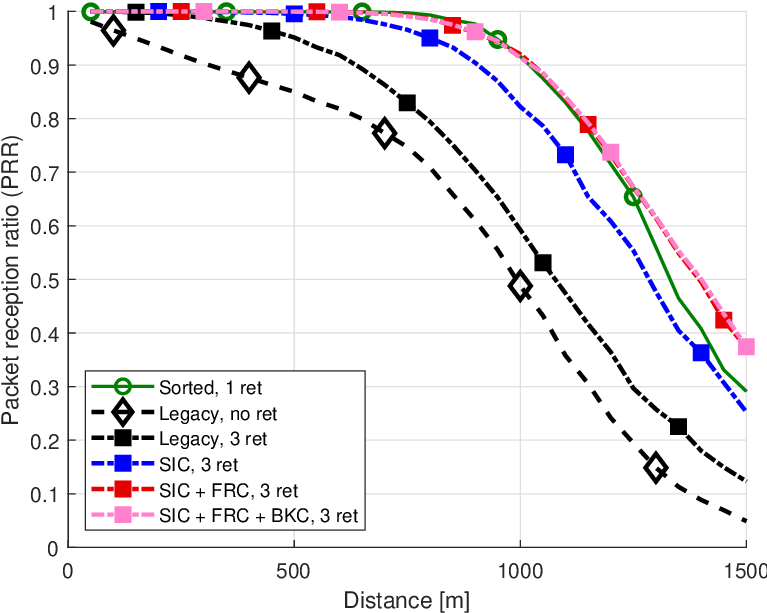}
         \caption{Aperiodic traffic, SB-DS.}
          \label{fig:MS_PRR_50_red_b}
     \end{subfigure} 
     \caption{PRR vs. transmitter-receiver distance, 12.5 veh/km.}
	\label{fig:MS_PRR_50_red}
\end{figure*}

\begin{figure*}[t] 
	\centering
      \begin{subfigure}[]{0.45\textwidth}
         \centering
         \includegraphics[width=0.9\columnwidth]{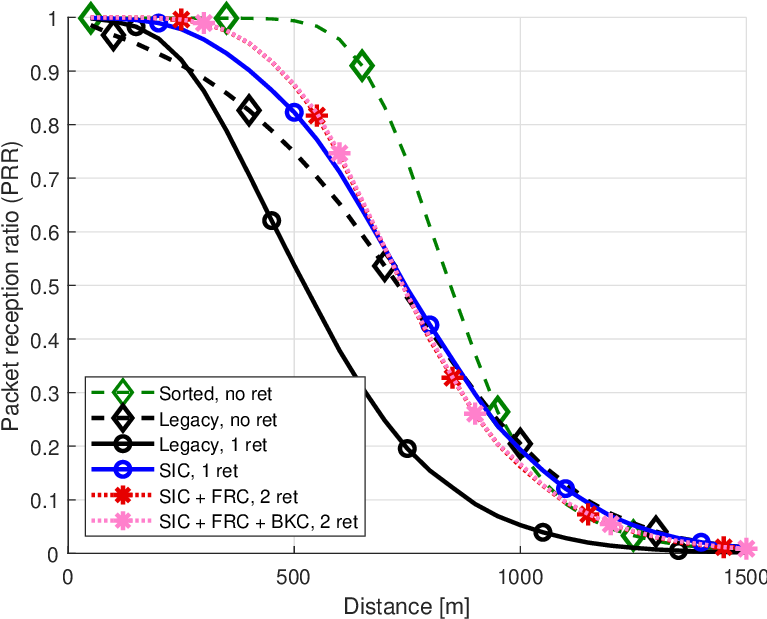}
         \caption{Periodic traffic, SB-SPS.}
          \label{fig:MS_PRR_200_red_a}
     \end{subfigure} 
      \begin{subfigure}[]{0.45\textwidth}
         \centering
         \includegraphics[width=0.9\columnwidth]{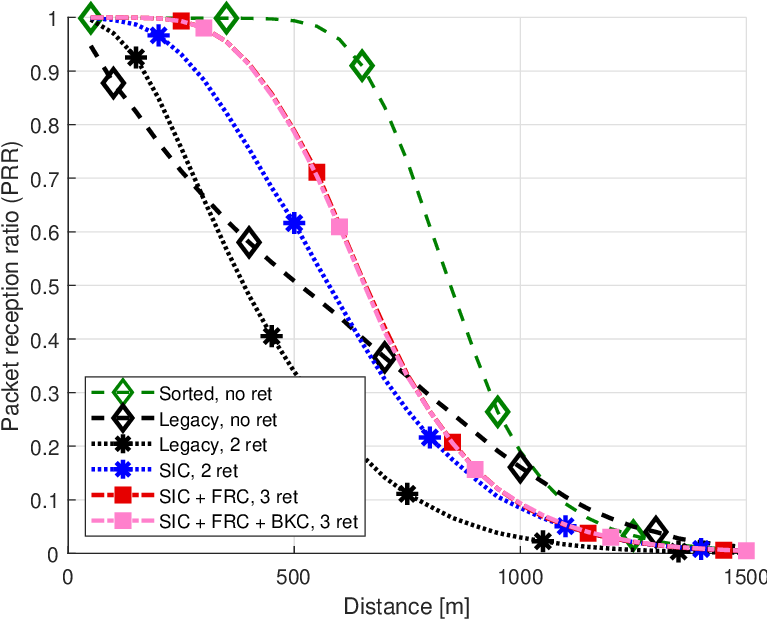}
         \caption{Aperiodic traffic, SB-DS.}
          \label{fig:MS_PRR_200_red_b}
     \end{subfigure} 
     \caption{PRR vs. transmitter-receiver distance, 50 veh/km.}
	\label{fig:MS_PRR_200_red}
\end{figure*}
Please remark that when SIC is foreseen, cancellation cannot anyway be applied if the strongest received signal cannot be decoded, i.e., if it is not sufficiently stronger than the interfering signals.

The aforementioned solutions are compared against an ideal resource allocation, denoted as \textit{sorted} \cite{8681439}. In that case, the vehicles are sorted based on their position and resources cyclically associated with the VUEs so that the distance between vehicles using the same resource is maximized and, as a consequence, the interference is minimized.  This process is continuously repeated. When $N$ copies of the same packet are transmitted, the resources are divided into $N$ groups and allocated one per copy, following the aforementioned principle. This implies that the average distance between \acp{VUE} using the same resources is reduced by a factor of $N$. An example of the sorted allocation is provided in Fig.~\ref{fig:ExampleSorted}, where four orthogonal resources are assumed.

\subsection{Results}
Results are reported  for periodic and aperiodic traffic, for various vehicle densities and number of blind retransmissions. In most cases, only a subset of all possible configurations for each solution is shown to improve the readability of the plots.

\subsubsection{Impact of retransmissions and data traffic generation on channel congestion} Fig.~\ref{fig:CBR_MS} reports the median of the \ac{CBR} when  varying the vehicle density. Given that the resource allocation process and the number of retransmissions do not depend on the type of receiver, the results in Fig.~\ref{fig:CBR_MS} are valid for all the compared schemes except for the sorted one. Not surprisingly, the higher the number of retransmissions and the vehicle density, the larger the CBR. The CBR is in general higher with periodic traffic and SB-SPS; this is expected because with SB-SPS the vehicles implement sensing procedures and, therefore, tend to select different resources when they are in the proximity of each other. It can also be observed that with 40 or more vehicles per km and more than one retransmission, the CBR is higher than 0.9, meaning high channel congestion. 

\begin{figure*}[h] 
	\centering
      \begin{subfigure}[]{0.45\textwidth}
         \centering
         \includegraphics[width=0.9\columnwidth]{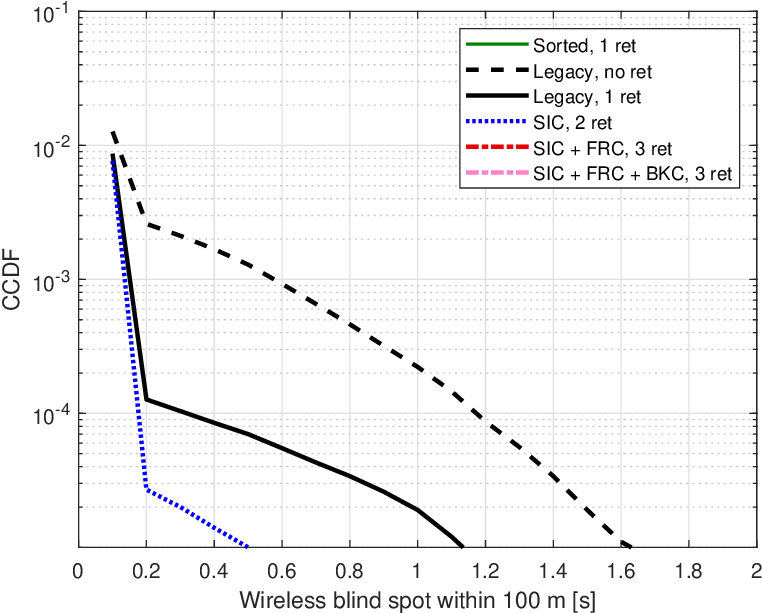}
         \caption{Periodic traffic, SB-SPS.}
          \label{fig:MS_WBS_50_red_a}
     \end{subfigure} 
      \begin{subfigure}[]{0.45\textwidth}
         \centering
         \includegraphics[width=0.9\columnwidth]{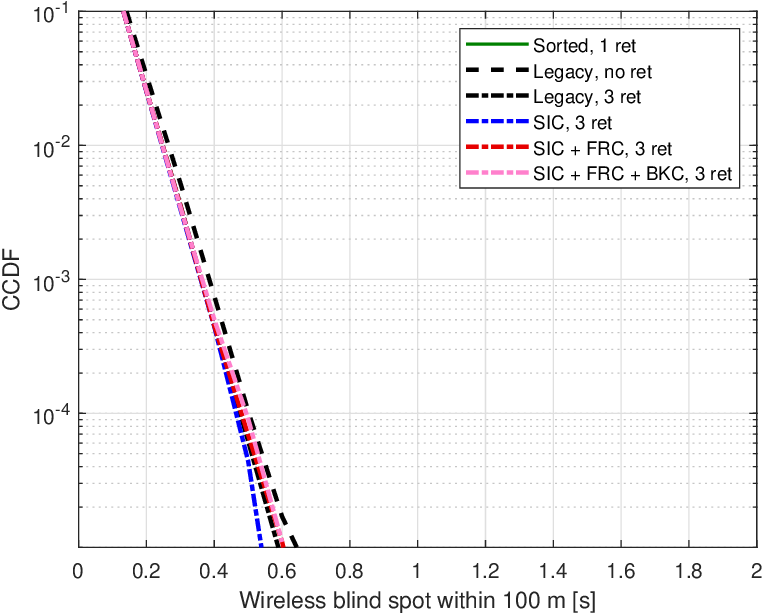}
         \caption{Aperiodic traffic, SB-DS.}
          \label{fig:MS_WBS_50_red_b}
     \end{subfigure} 
     \caption{CCDF of the WBSP, 12.5 veh/km.}
	\label{fig:MS_WBS_50_red}
\end{figure*}

\begin{figure*}[h] 
	\centering
      \begin{subfigure}[]{0.45\textwidth}
         \centering
         \includegraphics[width=0.9\columnwidth]{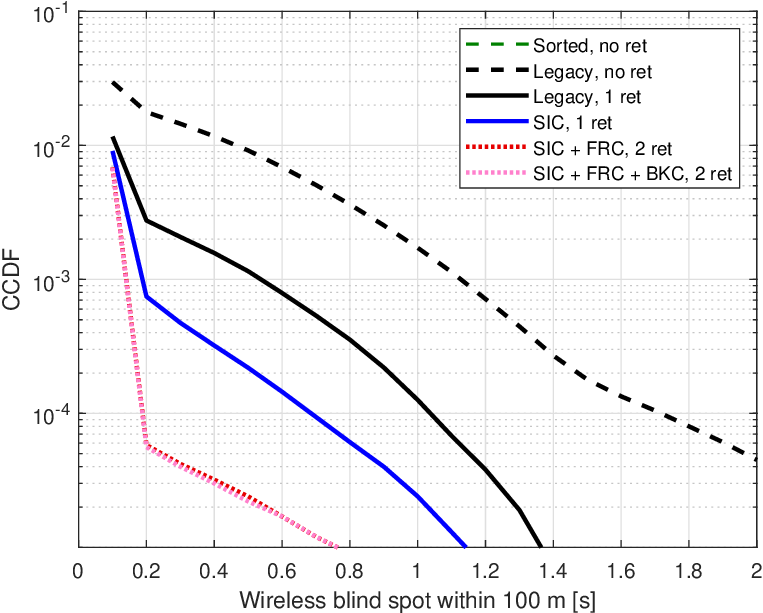}
         \caption{Periodic traffic, SB-SPS.}
          \label{fig:MS_WBS_200_red_a}
     \end{subfigure} 
      \begin{subfigure}[]{0.45\textwidth}
         \centering
         \includegraphics[width=0.9\columnwidth]{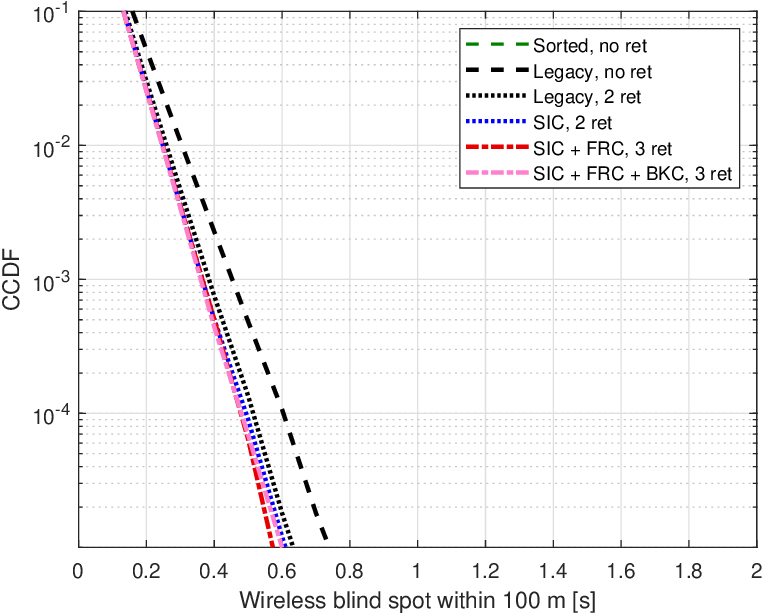}
         \caption{Aperiodic traffic, SB-DS.}
          \label{fig:MS_WBS_200_red_b}
     \end{subfigure} 
     \caption{CCDF of the WBSP, 50 veh/km.}
	\label{fig:MS_WBS_200_red}
\end{figure*}

\begin{figure}[t]
\centering
\includegraphics[width=0.85\columnwidth]{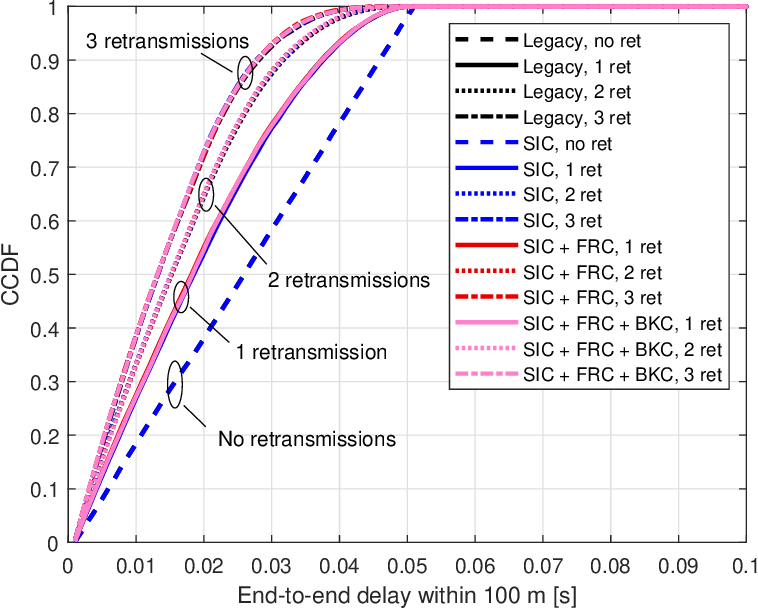}
\caption{CCDF of the EED, assuming periodic traffic and SB-SPS, with no to three retransmissions, 12.5 veh/km.}
\label{fig:EED}
\end{figure}

\subsubsection{Reliability analysis}

Figs.~\ref{fig:MS_PRR_50_red} and~\ref{fig:MS_PRR_200_red} show the PRR vs. the transmitter-receiver distance with a vehicle density of 12.5~veh/km and 50~veh/km, 
respectively, each including a subfigure focusing on periodic traffic (left) and another one  focusing on aperiodic traffic (right). In each subfigure and for each of the five solutions, only the curve corresponding to the number of retransmissions that maximizes the PRR values is shown; in addition, all subfigures include the legacy scheme without retransmissions as a worst case reference (dashed black curve). Looking at Fig.~\ref{fig:MS_PRR_50_red_a}, where periodic traffic is considered with low vehicle density (12.5~veh/km),  enforcing one retransmission in the legacy case provides a limited (but the highest) improvement when not using SIC. Indeed, under the legacy Mode 2, a number of retransmissions larger than one coupled with SB-SPS causes an excessive increase of congestion and, consequently, interference. Interestingly, all the proposed schemes instead approach the ideal sorted allocation (green curve).
Moving to the same density with aperiodic traffic (Fig.~\ref{fig:MS_PRR_50_red_b}), similar considerations are derived, although the difference from the worst to the best case and between cases amplifies. With aperiodic traffic and SB-DS, a number of retransmissions larger than one is beneficial also for the legacy Mode 2, which is consistent with results shown in \cite{campolo2021improving}. Also in this case, using SIC with \ac{FRC}, with or without \ac{BKC}, brings to results close to those obtained with the sorted allocation. 

\begin{figure*}[h] 
	\centering
      \begin{subfigure}[]{0.45\textwidth}
         \centering
         \includegraphics[width=0.9\columnwidth]{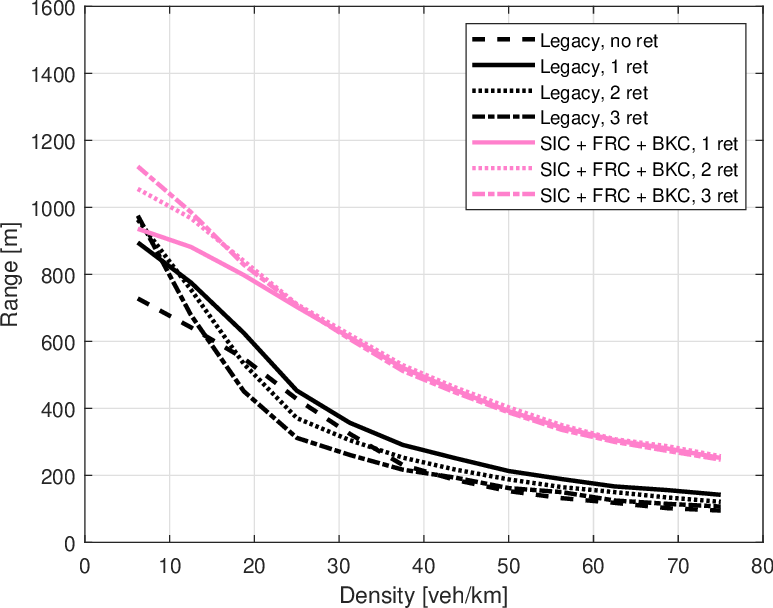}
         \caption{Periodic traffic, SB-SPS.}
          \label{fig:MS_Distance_red_a}
     \end{subfigure} 
      \begin{subfigure}[]{0.45\textwidth}
         \centering
         \includegraphics[width=0.9\columnwidth]{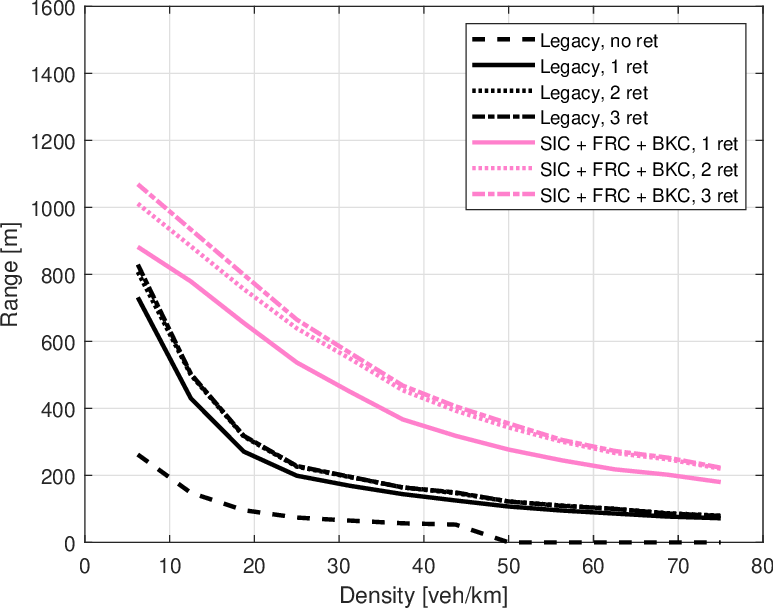}
         \caption{Aperiodic traffic, SB-DS.}
          \label{fig:MS_Distance_red_b}
     \end{subfigure} 
     \caption{Range vs. vehicle density, assuming legacy or SIC with \ac{FRC} and \ac{BKC}, with no to 3 retransmissions.} 
	\label{fig:MS_Distance_red}
\end{figure*}

\begin{figure}[t]
\centering
\includegraphics[width=0.85\columnwidth]{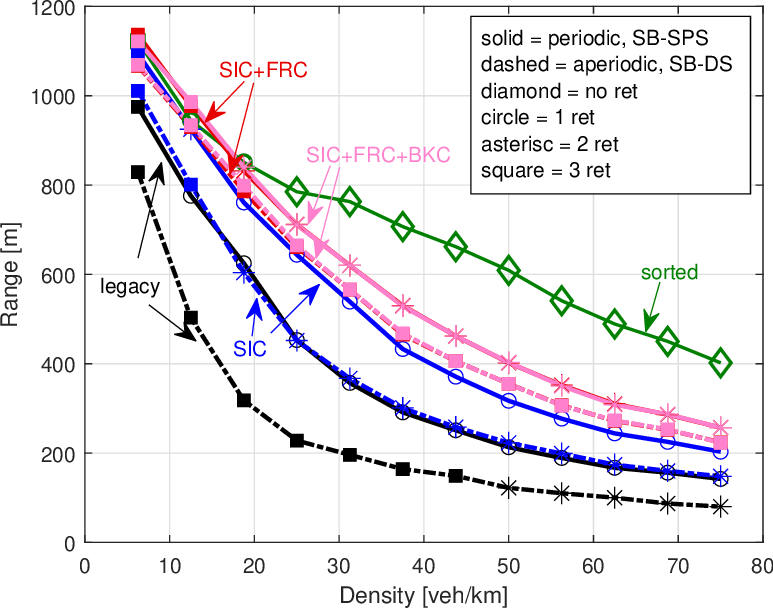}
\caption{Range vs. vehicle density, assuming for each density the best number of retransmissions. SCS 15 kHz and MCS 5.} 
\label{fig:MS_Distance_Per_vs_Aper}
\end{figure}

\begin{figure}[t]
\centering
\includegraphics[width=0.85\columnwidth]{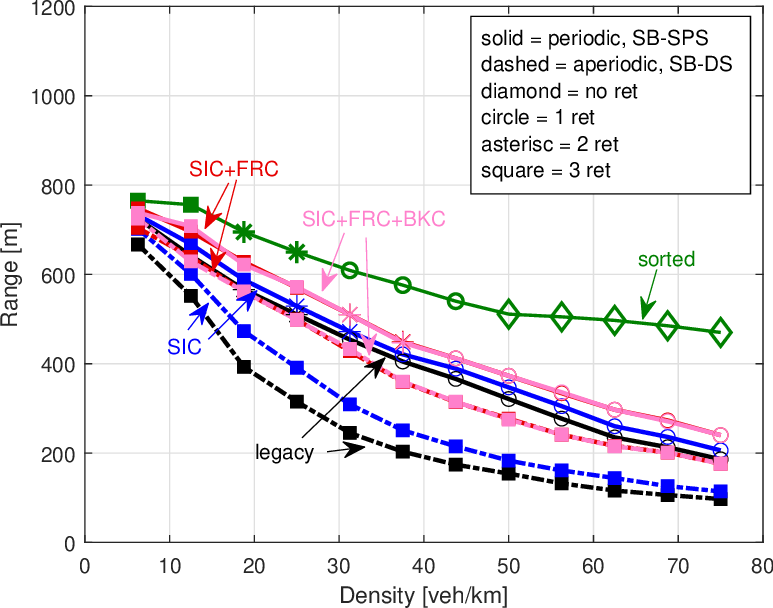}
\caption{Range vs. vehicle density, assuming for each density the best number of retransmissions. SCS 30~kHz and MCS~11.}
\label{fig:dist-range-SCS30}
\end{figure}

Observing Fig.~\ref{fig:MS_PRR_200_red}, where higher vehicle density is considered, the gap between what is achieved with legacy receivers and what is possible with the sorted allocation, is larger. Still, the scheme SIC and \ac{FRC} (with or without \ac{BKC}) outperforms the legacy Mode 2. Also the improvement from SIC to SIC+\ac{FRC} is significant, especially in the case of aperiodic traffic (Fig.~\ref{fig:MS_PRR_200_red_b}). In all the observed cases, the gain introduced by \ac{BKC} to SIC+FRC is very limited; albeit this is apparently a negative result, we can remark that: \textit{(i)} it was not known \textit{a priori} and, therefore, relevant from a scientific point of view; and \textit{(ii)} it means that the simpler solution consisting in the implementation of SIC+FRC is a good trade-off between performance and complexity.

\subsubsection{Wireless Blind Spot} Figs.~\ref{fig:MS_WBS_50_red} and~\ref{fig:MS_WBS_200_red} then show the \ac{CCDF} of the \ac{WBSP} for the same 12.5~veh/km and 50~veh/km  densities, respectively, considering the same cases shown in Figs.~\ref{fig:MS_PRR_50_red} and~\ref{fig:MS_PRR_200_red}. In all the figures, the curves related to the sorted allocation are not visible; in fact, the probability to have a \ac{WBS} within 100~m is lower than $10^{-5}$ also for the shortest considered interval, which is 100~ms. Comparing the other solutions, it is apparent that with periodic traffic the variability is much higher than with aperiodic traffic; this is expected because of the issue of persistent collisions that can occur with SB-SPS. It is also observable from Figs.~\ref{fig:MS_WBS_50_red} and~\ref{fig:MS_WBS_200_red} that the use of SIC strongly reduces the \ac{WBSP} compared to the legacy solution; \ac{FRC} further reduces the \ac{WBSP} in the case of periodic traffic, especially with small vehicle density, while the addition of \ac{BKC} impacts negligibly.

\subsubsection{Latency analysis}

Fig.~\ref{fig:EED} reports the \ac{CCDF} of the \ac{EED} when assuming 12.5 veh/km for the legacy, SIC, SIC+FRC, and SIC+FRC+BKC cases, with no to three retransmissions. What can be observed is that the latency slightly changes with the number of retransmissions, because by increasing the number of retransmissions inside the same interval $[T_1, T_2]$, the time when the first copy is correctly decoded statistically reduces. However, the curves corresponding to different cases but the same number of retransmissions almost overlap. As a minor remark, also recall that in all the cases the maximum possible access delay is bounded by the delay budget, which is set through the parameter $T_2$ (equal to 50~ms in our simulations).

\subsubsection{Range assessment}
Figs.~\ref{fig:MS_Distance_red} and~\ref{fig:MS_Distance_Per_vs_Aper} analyze the range when varying the vehicle density and the number of retransmissions from 1 to 3. Fig.~\ref{fig:MS_Distance_red}, in particular, compares the legacy and the complete RB-NOMA solutions with SIC+\ac{FRC}+\ac{BKC} mechanisms. As observable, the gain with SIC and \ac{FRC}+\ac{BKC} goes from a minimum of approximately 10\% (periodic traffic, very small vehicle density) to a maximum that exceeds 130\% (aperiodic traffic, high vehicle density). These results again remark that the impact of interference cancellation is higher with aperiodic traffic, where the allocation process is less effective in minimizing the interference. Two interesting additional observations are that: 1) the choice of the number of retransmissions is less critical with RB-NOMA than in the legacy scheme; and 2)  the difference in performance between periodic and aperiodic traffic is smaller with RB-NOMA than in the legacy scheme. 

Finally, Fig.~\ref{fig:MS_Distance_Per_vs_Aper} provides a summary, including all the solutions and both periodic and aperiodic traffic; for each vehicle density, the best number of retransmissions is assumed (i.e., the number of retransmissions that brings to the highest range). The markers in Fig.~\ref{fig:MS_Distance_Per_vs_Aper} indicate the number of retransmissions that are considered in each point of each curve. It is confirmed that the improvement from legacy to SIC and from SIC to SIC+\ac{FRC} is significant, whereas the improvement from SIC+\ac{FRC} to SIC+\ac{FRC}+\ac{BKC} is very limited. With low density (up to 20 veh/km), the use of RB-NOMA brings to a maximum distance that is very close to the (ideal) sorted allocation. Focusing on the markers: with the sorted allocation, the best is to have no retransmissions except for small vehicle density; with periodic traffic, it is preferable to use a smaller number of retransmissions (1 or 2 in most of the cases); with aperiodic traffic, it is preferable to use a larger number of retransmissions (3 in most of the cases). Finally, a key observation worth repeating is that, while in the legacy case there is a large difference between periodic traffic (solid curves) and aperiodic traffic (dashed curves), the difference is much smaller with RB-NOMA. This indicates that RB-NOMA effectively reduces the range gap between unpredictable (aperiodic) and predictable (periodic) traffic generation, addressing a known weakness of NR-V2X.

\subsubsection{Discussion with larger MCS and SCS} The achieved results have been verified when adopting different MCS, SCS, and bandwidth settings. Among the results obtained with the various configurations tested, not all inserted here for the sake of brevity, Fig.~\ref{fig:dist-range-SCS30} shows the range vs. density when assuming SCS~30~kHz and MCS~11. MCS is set to 11 as it is the lowest MCS to allocate with SCS~30~kHz one 1000~bytes-long packet in one TTI. MCS~11 corresponds to 4-QAM with $R_\text{c}=0.37$, and implies a SINR threshold of 9.9~dB. Adopting higher SCS and MCS allows on the one hand to double the number of packets that can be accommodated over the same time window (a single packet occupies a 0.5~ms-long TTI, instead of a 1~ms-long TTI), but on the other hand requires a higher SINR for successful decoding. As it can be observed from Fig.~\ref{fig:dist-range-SCS30}, the overall performance is worse when adopting 30~kHz and MCS~11, confirming the validity of our settings.

More importantly, what can be inferred from Fig.~\ref{fig:dist-range-SCS30} and was also observed with other settings, is that the overall conclusions that are derived by comparing the legacy, the SIC, and the SIC+FRC (with or without BKC) do not change.
 
\section{Conclusion}\label{sec:conclusions}

In this work, we proposed and analyzed an extension of the Mode 2 autonomous resource allocation foreseen by 5G NR-V2X, named RB-NOMA. According to the proposal, advanced receivers equipped with \ac{SIC} capabilities perform interference cancellation, based on the decoded messages and on the carried control information about previous and next copies that are sent blindly to improve the reliability. Extensive network-level simulations  were performed under a large number of different settings, showing the significant improvement compared to the use of legacy receivers. The gain in terms of range is up to approximately 70\% with periodic traffic and exceeds 130\% with aperiodic traffic, compared to the legacy Mode~2.

More in detail, the main observations deriving from the achieved results can be summarized as follows:
\begin{itemize}
    \item Similarly to what early presented in \cite{todisco2023sic}, SIC strongly improves the performance in terms of both PRR and WBSP; the use of the proposed forward cancellation (\ac{FRC}) further improves the performance and the improvement is remarkable with higher vehicle density and with aperiodic traffic, which are expected to be the most challenging (but likely) conditions for the sidelink;
    \item In all simulated scenarios, the use of backward cancellation (\ac{BKC}) negligibly improves the performance further when compared to the FRC; this is especially interesting noting that its implementation increases the complexity compared to SIC and to \ac{FRC} receivers;
    \item Minor modifications to the legacy 3GPP specifications for Mode 2 are needed to implement RB-NOMA; in particular, each transmission needs to carry indications about all future allocations with FRC and all past allocations with BKC, which may negligibly increase the control overhead for some of the transmissions;
    \item When the vehicle density is low, the use of RB-NOMA approaches very closely the ideal case; some margin for further improvement is still present for the large vehicle density case;
    \item Assuming RB-NOMA, the number of retransmissions have a lower impact than in the legacy case; this is helpful, because the optimal setting of the number of retransmissions may not be a trivial task;
    \item The improvement in all the cases appears more evident with aperiodic traffic; 
    \item When adopting RB-NOMA, the performance with periodic and aperiodic traffic is very close; moving the complexity from the resource allocation at the transmitter, which attempts to limit interference a priori, to the signal processing at the receiver, which reduces the impact of interference a posteriori, allows to reduce the importance of the packet generation pattern; this improves the independence between the access layer and the application. 
\end{itemize}

\section*{Acknowledgment}

This work has been conducted in the framework of the CNIT-WiLab and the WiLab-Huawei Joint Innovation Center.

\section*{Appendix A: Mathematical description of the SIC process}

This appendix describes mathematically the SIC process. Let us focus on a generic tagged \ac{VUE}, receiving the signal in a given \ac{TTI}. The neighboring \acp{VUE} transmitting in the same TTI are $\text{u}_1$, $\text{u}_2$, ..., $\text{u}_{|\Uset|}$, where $\Uset$ is the set of all \acp{VUE} transmitting in the neighborhood of the tagged vehicle in the considered time slot and $|\Uset|$ is the cardinality of the set $\Uset$. The samples of the signal received by the VUE under observation can be written as: 
\begin{equation}\label{eq:y_generic}
    y(k) = \sum_{j=1}^{|\Uset|} h_{\text{u}_j}(k) x_{\text{u}_j}(k) + n(k),
\end{equation}
where $k$ is used to indicate the sample at time $t_k$, $x_{\text{u}_j}$ is the signal received from the station $\text{u}_j$, $h_{\text{u}_j}$ is the channel coefficient from the station $\text{u}_j$ to the tagged VUE, and $n$ is the \ac{AWGN} contribution.

Assuming that the signal received from the station $u_1$ is the strongest signal, let us rewrite   \eqref{eq:y_generic} as:
\begin{align}\label{eq:y_generic2}
    y(k) &= h_{\text{u}_1}(k) x_{\text{u}_1}(k) + \sum_{j=2}^{|\Uset|} h_{\text{u}_j}(k) x_{\text{u}_j}(k) + n(k)\nonumber\\ &= h_{\text{u}_1}(k) x_{\text{u}_1}(k) + \vartheta(k) + n(k),
\end{align}
where $\vartheta(k) \triangleq \sum_{j=2}^{|\Uset|} h_{\text{u}_j}(k) x_{\text{u}_j}(k)$ is the contribution from the interference. If the receiver detects a signal, it is the signal from $u_1$; in such case, it tries to decode the data it carries. If the detection succeeds, the tagged VUE is able to regenerate perfectly the signal $x_{\text{u}_1}(k)$. Starting from the exact transmitted signal, the VUE can use it to improve the estimation of the channel $h_{\text{u}_1}(k)$; since all the bits are known, in fact, all the subcarriers are used as they were carrying pilot signals. Then, denoting the estimated channel as $\hat{h}_{\text{u}_1}(k)$, the receiver can subtract from $y$ the estimated signal received from user $\text{u}_1$ and obtain a new signal $y_{I1}$, which can be written as:
\begin{align}\label{eq:y_I1}
    y_{I1}(k) &= y(k) - \hat{h}_{\text{u}_1}(k) x_{\text{u}_1}(k) \\ &= \left( h_{\text{u}_1}(k) - \hat{h}_{\text{u}_1}(k)\right) x_{\text{u}_1}(k) + \sum_{j=2}^{|\Uset|} h_{\text{u}_j}(k) x_{\text{u}_j}(k) + n(k)\;. \nonumber
\end{align}

We can rewrite \eqref{eq:y_I1} as:
\begin{align}\label{eq:y_I1_bis}
    y_{I1}(k) &= h_{\text{u}_2}(k) x_{\text{u}_2}(k) + \left(h_{\text{u}_1}(k)-\hat{h}_{\text{u}_1}(k)\right) x_{\text{u}_1}(k) \nonumber \\
    &+ \sum_{j=3}^{|\Uset|} h_{\text{u}_j}(k) x_{\text{u}_j}(k) + n(k) \nonumber \\
    &= h_{\text{u}_2}(k) x_{\text{u}_2}(k) + \vartheta_{I1}(k) + n(k),
\end{align}
where $\vartheta_{I1}(k) \triangleq \left(h_{\text{u}_1}(k)-\hat{h}_{\text{u}_1}(k)\right) x_{\text{u}_1}(k) + \sum_{j=3}^{|\Uset|} h_{\text{u}_j}(k) x_{\text{u}_j}(k)$ represents the contribution from the residual interference.
If the signal received from $\text{u}_2$ is sufficiently stronger than the residual interference plus noise, then the receiver can detect the signal from $u_2$ and decode the data it carries.

The process can be reiterated until the strongest remaining contribution is not detectable or the data cannot be decoded. In our results, we limit the process to at most one single cancellation to avoid introducing excessive processing delay.

\section*{Appendix B: implementation details} In the simulator, \ac{FRC} and \ac{BKC} are implemented through the following recursive procedure:
\begin{enumerate}
    \item The TTI elaborated, indicated with $t$ is set to the current TTI $t_0$, and the process goes to step~2;
    \item The values of $\xi_{i,\text{u}_j}$, with $\text{u}_j \in \Uset_i$ are updated; Denoting with $\Uset^{\#}_i \subset \Uset_i$ the set of transmitting neighbors in TTI $t$ for which $i$ has not yet decoded a copy of the packet, it is $\xi_{i,\text{u}_j}=0$ if $\text{u}_j \in \Uset^{\#}_i$ and $\xi_{i,\text{u}_j}=1$ otherwise;
    \item The receiver attempts to decode $\text{u}_m \in \Uset^{\#}_i$ corresponding to the minimum index $m$ (i.e., it tries to decode the packet from the strongest source among those that have a packet not yet decoded in the TTI); this is done calculating \eqref{NOMAgenericSINR3} using $\text{u}_m$ instead of $\text{u}_z$, and then comparing the resulting SINR with the SINR threshold $\SINRt$;
    \item If the packet from user $\text{u}_m$ is not decoded (i.e., if $\gamma_{i,\text{u}_m}<\SINRt$), then this process ends;
    \item If the packet from user $\text{u}_m$ is decoded (i.e., if $\gamma_{i,\text{u}_m} \geq \SINRt$), the following additional operations are performed, some of which only hold in the case that also \ac{BKC} is considered: 
    \begin{itemize}
        \item First, only if also \ac{BKC} is used, if a copy of the decoded packet is indicated in the past TTI $t_{-n}$, then $t$ is set to $t_{-n}$ and a new process is started from step~2;
        \item Second, only if also \ac{BKC} is used, if a copy of the decoded packet is indicated in the future TTI $t_{m}$ and $t_{m}\leq t_0$, then $t$ is set to $t_{m}$ and a new process is started from step~2;
        \item Lastly, also if only \ac{FRC}, the $\xi_{i,\text{u}_m}$ is updated to 1 and a new process is started from step~3.
    \end{itemize}
\end{enumerate}

\bibliographystyle{IEEEtran}\bibliography{biblio,myBibOnlyRelated}

@inproceedings{campolo2021improving,
  title={Improving Resource Allocation for beyond {5G V2X} Sidelink Connectivity},
  author={Campolo, Claudia and Todisco, Vittorio and Molinaro, Antonella and Berthet, Antoine and Bartoletti, Stefania and Bazzi, Alessandro},
  booktitle={2021 55th Asilomar Conference on Signals, Systems, and Computers},
  pages={55--60},
  year={2021},
  organization={IEEE}
}

@ARTICLE{8681439,
  author={Bazzi, Alessandro and Zanella, Alberto and Cecchini, Giammarco and Masini, Barbara M.},
  journal={IEEE Transactions on Vehicular Technology}, 
  title={Analytical Investigation of Two Benchmark Resource Allocation Algorithms for {LTE-V2V}}, 
  year={2019},
  volume={68},
  number={6},
  pages={5904-5916},
  doi={10.1109/TVT.2019.2909438}}

@article{todisco2021performance,
  title={{Performance analysis of sidelink 5G-V2X mode 2 through an open-source simulator}},
  author={Todisco, Vittorio and Bartoletti, Stefania and Campolo, Claudia and Molinaro, Antonella and Berthet, Antoine O and Bazzi, Alessandro},
  journal={IEEE Access},
  volume={9},
  pages={145648--145661},
  year={2021},
  publisher={IEEE}
}

@ARTICLE{10078378,
  author={Bazzi, Alessandro and Campolo, Claudia and Todisco, Vittorio and Bartoletti, Stefania and Decarli, Nicolò and Molinaro, Antonella and Berthet, Antoine O. and Stirling-Gallacher, Richard A.},
  journal={IEEE Vehicular Technology Magazine}, 
  title={Toward {6G} Vehicle-to-Everything Sidelink: Nonorthogonal Multiple Access in the Autonomous Mode}, 
  year={2023},
  volume={18},
  number={2},
  pages={50-59},
  doi={10.1109/MVT.2023.3252278}}

@ARTICLE{ETSI_103_574, 
	author = {ETSI},
journal={ETSI TS 103 574 V1.1.1}, 
	title={Intelligent Transport Systems {(ITS)}; Congestion Control Mechanisms for the {C-V2X PC5} interface;
	Access layer part}, 
	year={2018}, 
	month={Nov.},}

@article{bazzi2020wireless,
  title={{On wireless blind spots in the C-V2X sidelink}},
  author={Bazzi, Alessandro and Campolo, Claudia and Molinaro, Antonella and Berthet, Antoine O and Masini, Barbara M and Zanella, Alberto},
  journal={IEEE Transactions on Vehicular Technology},
  volume={69},
  number={8},
  pages={9239--9243},
  year={2020},
  publisher={IEEE}
}

@misc{3GPP_TR_36_885,
  author = "3GPP",
  title={{TR} 36.885 V16.2.0. {T}echnical Specification Group Radio Access Network;
	{S}tudy on {LTE}-based {V2X} services},
  year={2019},
  month={July}
}

@misc{c2c-2020_spectrum, 
	author={}, 
	journal={C2C-CC position paper}, 
	title={Position Paper on Road Safety and Road Efficiency Spectrum Needs in the 5.9 {GHz} for {C-ITS} and Cooperative Automated Driving}, 
	organization  = {C2C-CC},
	year={2020}
}

@article{novak2013idma,
  title={{IDMA for the multiuser MIMO-OFDM uplink: A factor graph framework for joint data detection and channel estimation}},
  author={Novak, Clemens and Matz, Gerald and Hlawatsch, Franz},
  journal={IEEE Transactions on Signal Processing},
  volume={61},
  number={16},
  pages={4051--4066},
  year={2013},
  publisher={IEEE}
}

@inproceedings{rajalakshmi2024enhancing,
  title={{Enhancing Sidelink V2X Communication in 6G Networks: Power-Domain NOMA-Based Priority Message Transmission Approach}},
  author={Rajalakshmi, P and others},
  booktitle={2024 IEEE Vehicular Networking Conference (VNC)},
  pages={117--124},
  year={2024},
  organization={IEEE}
}

@misc{3GPP37885,
  author = "3GPP",
  title={{TR} 37.885 v15.3.0, {Study on evaluation methodology of new Vehicle-to Everything {(V2X)} use cases for {LTE} and {NR}}; {R}elease 15},
  year={2019},
  month={June},
  institution={3rd Generation Partnership Project},
}

@misc{3GPP_TR_22_885,
  author = "3GPP",
  title={{TR} 22.885 V14.0.0. {S}tudy on {LTE} Support for {V2X} Services},
  year={2015},
  month={December}
}

@inproceedings{tweed2018dynamic,
  title={Dynamic resource allocation for uplink {MIMO NOMA VWN} with imperfect {SIC}},
  author={Tweed, Daniel and Le-Ngoc, Tho},
  booktitle={2018 IEEE ICC},
  pages={1--6}
}

@Article{WuBarMarBaz:J22,
AUTHOR = {Wu, Zhuofei and Bartoletti, Stefania and Martinez, Vincent and Bazzi, Alessandro},
TITLE = {A Methodology for Abstracting the Physical Layer of Direct {V2X} Communications Technologies},
JOURNAL = {Sensors},
VOLUME = {22},
YEAR = {2022},
NUMBER = {23},
ARTICLE-NUMBER = {9330}
}

@article{todisco2023sic,
  title={On the Performance of {SIC}-based {NOMA} in the {C-V2X} Sidelink Autonomous Mode},
  author={Todisco, Vittorio and Campolo, Claudia and Molinaro, Antonella and Berthet, Antoine O. and Stirling-Gallacher, Richard A. and Bazzi, Alessandro},
  journal={IEEE Conference on Standards for Communications and Networking (CSCN)},
  year={2023}
}

@article{bazzi2024mco,
  title={Multi-Channel Operation for the Release 2 of {ETSI} Cooperative Intelligent Transport Systems
Vehicular Networking},
  author={Bazzi, Alessandro and Sepulcre,  Miguel and Delooz, Quentin and Festag, Andreas  and Vogt, Jonas  and Wieker, Horst and Berens, Friedbert  and Spaanderman, Paul},
  journal={IEEE Communications Standards Magazine},
  year={2024}
}

@ARTICLE{9762711,
  author={Delooz, Quentin and Willecke, Alexander and Garlichs, Keno and Hagau, Andreas-Christian and Wolf, Lars and Vinel, Alexey and Festag, Andreas},
  journal={IEEE Access}, 
  title={Analysis and Evaluation of Information Redundancy Mitigation for {V2X} Collective Perception}, 
  year={2022},
  volume={10},
  number={},
  pages={47076-47093},
  doi={10.1109/ACCESS.2022.3170029}}

@ARTICLE{10195908,
  author={Molina-Masegosa, Rafael and Avedisov, Sergei S. and Sepulcre, Miguel and Farid, Yashar Z. and Gozalvez, Javier and Altintas, Onur},
  journal={IEEE Vehicular Technology Magazine}, 
  title={{V2X} Communications for Maneuver Coordination in Connected Automated Driving: Message Generation Rules}, 
  year={2023},
  volume={18},
  number={3},
  pages={91-100},
  doi={10.1109/MVT.2023.3284562}}

@ARTICLE{8357810,
  author={Dai, Linglong and Wang, Bichai and Ding, Zhiguo and Wang, Zhaocheng and Chen, Sheng and Hanzo, Lajos},
  journal={IEEE Communications Surveys \& Tutorials}, 
  title={A Survey of Non-Orthogonal Multiple Access for {5G}}, 
  year={2018},
  volume={20},
  number={3},
  pages={2294-2323},
  doi={10.1109/COMST.2018.2835558}}

@ARTICLE{7676258,
  author={Islam, S. M. Riazul and Avazov, Nurilla and Dobre, Octavia A. and Kwak, Kyung-sup},
  journal={IEEE Communications Surveys \& Tutorials}, 
  title={Power-Domain Non-Orthogonal Multiple Access {(NOMA)} in {5G} Systems: Potentials and Challenges}, 
  year={2017},
  volume={19},
  number={2},
  pages={721-742},
  doi={10.1109/COMST.2016.2621116}}

@INPROCEEDINGS{9287485,
  author={Tahir, Bashar and Schwarz, Stefan and Rupp, Markus},
  booktitle={2020 28th European Signal Processing Conference (EUSIPCO)}, 
  title={Collision Resilient {V2X} Communication via Grant-Free {NOMA}}, 
  year={2021},
  volume={},
  number={},
  pages={1732-1736},
  doi={10.23919/Eusipco47968.2020.9287485}}

@ARTICLE{7302046,
  author={Paolini, Enrico and Liva, Gianluigi and Chiani, Marco},
  journal={IEEE Transactions on Information Theory}, 
  title={Coded Slotted {ALOHA}: A Graph-Based Method for Uncoordinated Multiple Access}, 
  year={2015},
  volume={61},
  number={12},
  pages={6815-6832},
  doi={10.1109/TIT.2015.2492579}}

@inproceedings{saito2013non,
  title={{Non-orthogonal multiple access (NOMA) for cellular future radio access}},
  author={Saito, Yuya and Kishiyama, Yoshihisa and Benjebbour, Anass and Nakamura, Takehiro and Li, Anxin and Higuchi, Kenichi},
  booktitle={2013 IEEE 77th vehicular technology conference (VTC Spring)},
  pages={1--5},
  year={2013},
  organization={IEEE}
}

@inproceedings{manglayev2017noma,
  title={{NOMA with imperfect SIC implementation}},
  author={Manglayev, Talgat and Kizilirmak, Refik Caglar and Kho, Yau Hee and Bazhayev, Nurzhan and Lebedev, Ilya},
  booktitle={IEEE EUROCON 2017-17th International Conference on Smart Technologies},
  pages={22--25},
  year={2017},
  organization={IEEE}
}

@inproceedings{almohamad2020low,
  title={{Low complexity constellation rotation-based SIC detection for IM-NOMA schemes}},
  author={Almohamad, Abdullateef and Hasna, Mazen and Althunibat, Saud and {\"O}zyurt, Serdar and Qaraqe, Khalid},
  booktitle={2020 IEEE 92nd Vehicular Technology Conference (VTC2020-Fall)},
  pages={1--5},
  year={2020},
  organization={IEEE}
}

@article{xu2021low,
  title={{A Low-Latency Random Access Scheme by Multichannel SIC for Industrial IoT}},
  author={Xu, Chaonong and Jiang, Biao and Li, Chao},
  journal={IEEE Systems Journal},
  volume={16},
  number={1},
  pages={810--819},
  year={2021},
  publisher={IEEE}
}

@article{sunkaraboina2024fpga,
  title={{FPGA-based hardware accelerator for SIC in uplink NOMA networks}},
  author={Sunkaraboina, Sreenu and Naidu, Kalpana},
  journal={Telecommunication Systems},
  volume={86},
  number={2},
  pages={383--392},
  year={2024},
  publisher={Springer}
}

@ARTICLE{etsi202110,
author={ETSI}, 
journal={{ETSI} {TR} 103 439-v2.1.1},
title={Intelligent Transport Systems {(ITS)}; Multi-Channel Operation study; Release 2},
year={2021},
}

@misc{fcc,
    institution = {Federal Communications Commission ({FCC})},
    note = {Version 1.0, \url{https://docs.fcc.gov/public/attachments/FCC-24-123A1.pdf}, Accessed: \today},
    title = {Second Report and
Order, {FCC}-20-164, Docket/RM: 19-138},
    year = {2024},
    month = {Nov.}
}

@article{daw2024lamp,
  title={{LAMP: A latency-aware MAC protocol for joint scheduling of CAM and DENM traffic over 5G-NR sidelink}},
  author={Daw, Suranjan and Kar, Anwesha and Chintapalli, Venkatarami Reddy and Tamma, Bheemarjuna Reddy and others},
  journal={Computer Communications},
  volume={217},
  pages={41--56},
  year={2024},
  publisher={Elsevier}
}

@misc{38.212,
  author = "3GPP",
  title       = "{{TS} 38.212 v18.6.0, Multiplexing and channel coding}",
  month       = "March",
  year        = "2024",
}

@misc{38.214,
  author = "3GPP",
  title       = "{{TS} 38.214 v18.6.0, Physical layer procedures for data}",
  month       = "March",
  year        = "2024",
}

@misc{38.101-1,
  author = "3GPP",
  title       = "{{TS} 38.101-1 v18.9.0, User Equipment (UE) radio transmission and reception; Part 1: Range 1 Standalone}",
  month     = "April",
  year        = "2024",
}

@article{di2017,
  title={Non-orthogonal multiple access for high-reliable and low-latency {V2X} communications in 5{G} systems},
  author={Di, Boya and Song, Lingyang and Li, Yonghui and Li, Geoffrey Ye},
  journal={IEEE journal on selected areas in communications},
  volume={35},
  number={10},
  pages={2383--2397},
  year={2017},
  publisher={IEEE}
}

@ARTICLE{casini2007,
  author={E. {Casini} and R. {De Gaudenzi} and O. {Del Rio Herrero}},
  journal={IEEE Transactions on Wireless Communications}, 
  title={Contention Resolution Diversity Slotted {ALOHA (CRDSA)}: An Enhanced Random Access Scheme for Satellite Access Packet Networks}, 
  year={2007},
  volume={6},
  number={4},
  pages={1408-1419},}

@ARTICLE{liva2011,
  author={G. {Liva}},
  journal={IEEE Transactions on Communications}, 
  title={Graph-Based Analysis and Optimization of Contention Resolution Diversity Slotted {ALOHA}}, 
  year={2011},
  volume={59},
  number={2},
  pages={477-487},}

@INPROCEEDINGS{ivanov2015,
  author={M. {Ivanov} and F. {Brännström} and A. G. i. {Amat} and P. {Popovski}},
  booktitle={2015 IEEE International Conference on Communication Workshop (ICCW)}, 
  title={All-to-all broadcast for vehicular networks based on coded slotted {ALOHA}}, 
  year={2015},
  volume={},
  number={},
  pages={2046-2050},}

@ARTICLE{ivanov2017,
  author={M. {Ivanov} and F. {Brännström} and A. {Graell i Amat} and P. {Popovski}},
  journal={IEEE Transactions on Communications}, 
  title={Broadcast Coded Slotted {ALOHA}: A Finite Frame Length Analysis}, 
  year={2017},
  volume={65},
  number={2},
  pages={651-662},}

@inproceedings{hirai2019,
  title={{NOMA} Concept for {PC}5-Based Cellular-{V2X} Mode 4 in Crash Warning System},
  author={Hirai, Takeshi and Murase, Tutomu},
  booktitle={2019 IEEE 90th Vehicular Technology Conference (VTC2019-Fall)},
  pages={1--6},
  year={2019},
  organization={IEEE}
}

@ARTICLE{GarMolBobGozColSahKou:21,
  author={M. H. C. {Garcia} and A. {Molina-Galan} and M. {Boban} and J. {Gozalvez} and B. {Coll-Perales} and T. {Şahin} and A. {Kousaridas}},
  journal={{IEEE} Communications Surveys   Tutorials}, 
  title={A Tutorial on {5G NR V2X} Communications}, 
  year={2021},
  pages={1-1}
  }

@article{bazzi2021design,
  title={On the Design of Sidelink for Cellular {V2X}: A Literature Review and Outlook for Future},
  author={Bazzi, Alessandro and Berthet, Antoine O and Campolo, Claudia and Masini, Barbara Mav{\`\i} and Molinaro, Antonella and Zanella, Alberto},
  journal={IEEE Access},
 year={2021},
  volume={9},
  pages={97953-97980},
}

\begin{IEEEbiography}[{\includegraphics[width=1in,height=1.2in,clip,keepaspectratio]{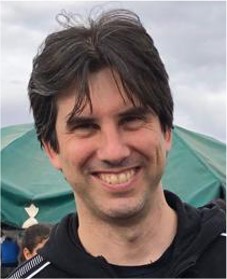}}]{Alessandro Bazzi}~(Senior Member, IEEE)  
is an Associate Professor at the University of Bologna. 
His research interests are mainly on medium access control and radio resource management of wireless networks, with particular focus on connected and autonomous vehicles (CAVs). He 
is/had been part of the ETSI Specialist Task Forces working on multi-channel and radio resource management in the ITS band. 
\end{IEEEbiography}

\begin{IEEEbiography}[{\includegraphics[width=1in,height=1.2in,clip,keepaspectratio]{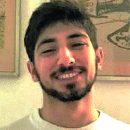}}]{Vittorio Todisco~(Student Member, IEEE)}
is a Postdoctoral Researcher at the University of Bologna. His research focuses on wireless systems for intelligent transportation, with particular emphasis on channel access and resource management. He also follows standardization activities related to vehicular communications, especially developments in the 3GPP NR-V2X technology.
\end{IEEEbiography}

\begin{IEEEbiography}[{\includegraphics[width=1in,height=1.2in,clip,keepaspectratio]{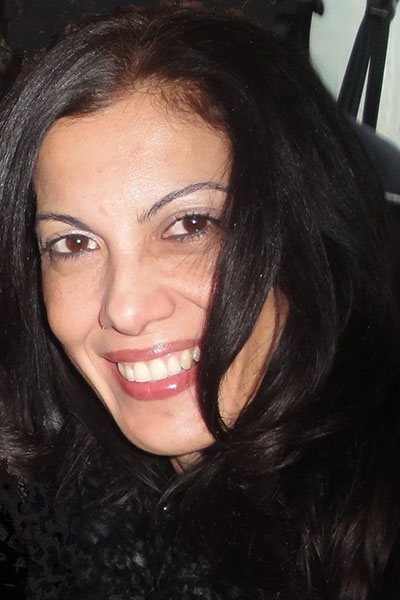}}]{Antonella Molinaro}~(Senior Member, IEEE)  is a full professor of telecommunications with the University Mediterranea of Reggio Calabria, Italy, and with CentraleSupélec – Université Paris-Saclay, France. Her research mainly focuses on wireless and mobile networking, vehicular networks, and future internet.
\end{IEEEbiography}

\begin{IEEEbiography}[{\includegraphics[width=1in,height=1.2in,clip,keepaspectratio]{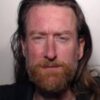}}]{Antoine O. Berthet}~(Senior Member, IEEE)
is a Full Professor at CentraleSupélec – Université Paris-Saclay in Gif-sur-Yvette, France, and a researcher at the Laboratoire des Signaux et Systèmes (CNRS UMR 8506). His research interests include information theory, communication theory, and signal processing, with a particular focus on enhancing the physical and access layers of satellite, cellular, and vehicular ad hoc networks.
\end{IEEEbiography}


\begin{IEEEbiography}
[{\includegraphics[width=1in,height=1.2in,clip,keepaspectratio]{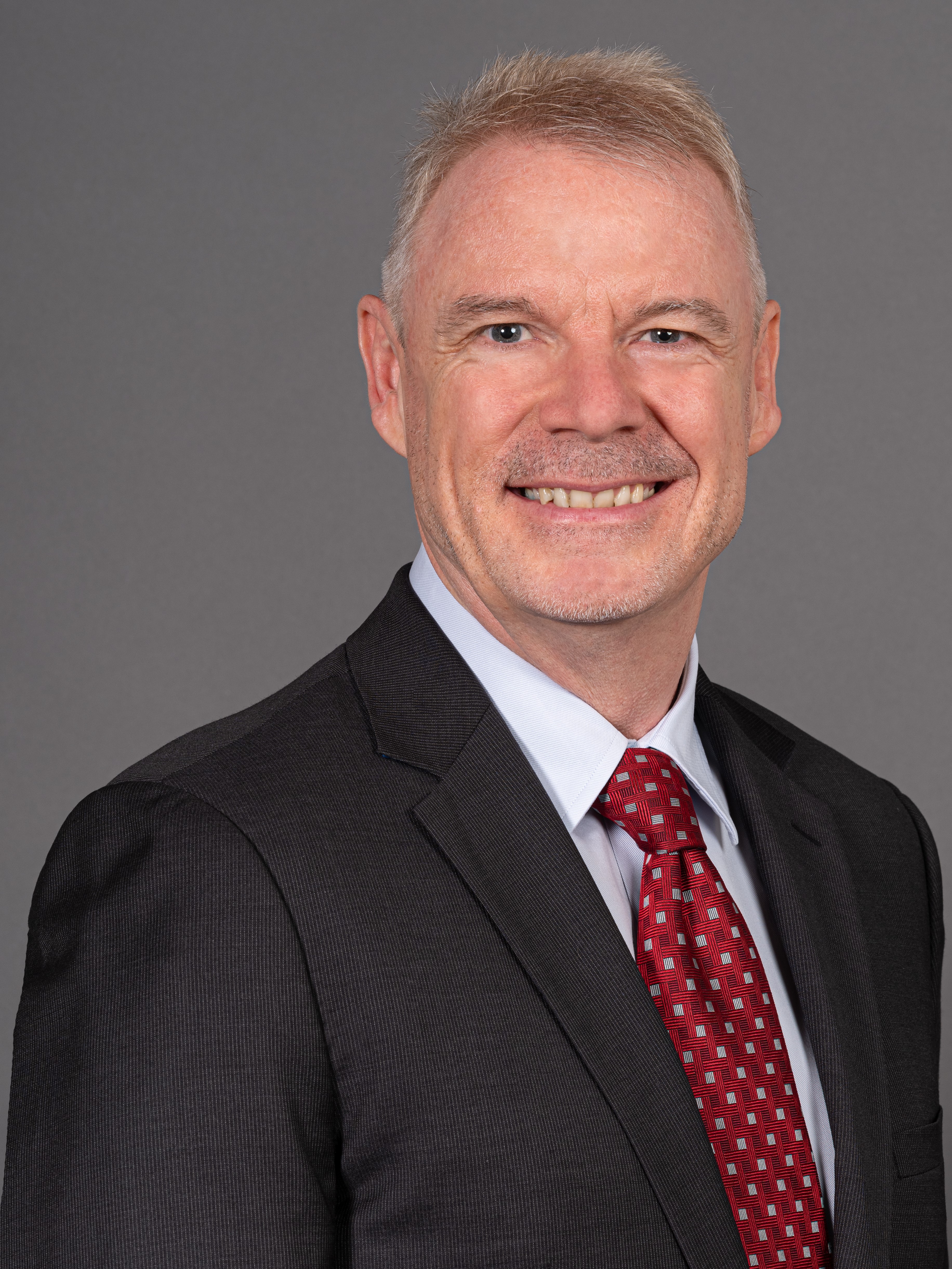}}]{Richard A. Stirling-Gallacher} (Member, IEEE) is a Research Expert/Team Leader of the Munich Research Center, Huawei Technologies Duesseldorf GmbH, Munich, Germany. He currently holds more than 140 granted U.S. patents and serves as vice chair for the ETSI ISG on ISAC. His current interests include ISAC, positioning, massive MIMO, and V2X for 6G communication systems.
\end{IEEEbiography}

\begin{IEEEbiography}[{\includegraphics[width=1in,height=1.2in,clip,keepaspectratio]{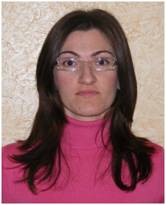}}]{Claudia Campolo}
is an associate professor of telecommunications at the University Mediterranea of Reggio Calabria. She received an master degree (2007) and a Ph.D. degree (2011) in telecommunications engineering  from the same university. Her main research interests are in the field of vehicular networking, future Internet architectures, 5G and beyond systems.
\end{IEEEbiography}

\end{document}